\documentclass[aps,pra,twocolumn,showpacs,superscriptaddress]{revtex4-1}
\usepackage[colorlinks,linkcolor=blue,anchorcolor=blue,citecolor=blue]{hyperref}
\usepackage{graphicx,amsmath,amssymb}
\usepackage[usenames]{color}
\bibpunct{[}{]}{,}{s}{}{;}

\begin{document}

\title{Scaling Relations and Topological Quadruple Points in Light-matter Interactions with Anisotropy and Nonlinear Stark Coupling
}

\author{Zu-Jian Ying }
\email{yingzj@lzu.edu.cn}
\affiliation{School of Physical Science and Technology, Lanzhou University, Lanzhou 730000, China}

\begin{abstract}
Universality is a common quality in different physical parameters that is rooted in the deep nature of physical systems. Scaling relation is a typical universality for critical phenomena around a quantum phase transition, while topological classification provides another type of universality essentially different from the critical universality. Both classes of universalities can be present in a single-qubit system with light-matter interactions, as exhibiting generally in the fundamental quantum Rabi model with anisotropy not only for linear coupling but also for nonlinear Stark coupling (NSC). In low frequencies different levels of scaling relations are demonstrated, holding for anisotropic or/and NSCs, locally or globally. At finite frequencies such a critical universality breaks down and diversity is dominant. However, common topological feature of the ground state can be extracted from the node number, which yields a topological class of universality amidst the critical diversity. Both conventional and unconventional topological transitions emerge, with their meeting, which never occurs in linear interaction, enabled by the nonlinear coupling to form topological quadruple points which are found to be spin-invariant points. Sensitivity analysis indicates that the NSC can be another approach to manipulate topological transitions in addition to coupling anisotropy.
\end{abstract}
\pacs{ }
\maketitle


\section{Introduction}

In the frontiers of modern quantum physics and quantum technologies, the
past decade has seen the great wave evoked by the extraordinary experimental
progresses \cite{Diaz2019RevModPhy,Kockum2019NRP} and tremendous theoretical
efforts \cite{Braak2011,Boite2020,Liu2021AQT} on the studies of light-matter
interactions. The remarkable realization of the ultra-strong\cite%
{Diaz2019RevModPhy,Wallraff2004,Gunter2009,Niemczyk2010,Peropadre2010,FornDiaz2017, Forn-Diaz2010,Scalari2012,Xiang2013,Yoshihara2017NatPhys,Kockum2017}
and even deep-strong couplings,\cite%
{Yoshihara2017NatPhys,Bayer2017DeepStrong} has opened the gate to a new
regime with a rich phenomenology unexpected in weak couplings. On the other
hand, the milestone work of D. Braak revealing the integrability \cite{Braak2011} of the
quantum Rabi model (QRM),\cite{rabi1936}  which is a most fundamental model
of light-interactions, has triggered an intensive dialogue between
mathematics and physics~\cite{Solano2011} and leads to a boom of theoretical
developments.~\cite%
{Boite2020,Wolf2012,FelicettiPRL2020,Felicetti2018-mixed-TPP-SPP,Felicetti2015-TwoPhotonProcess,Simone2018,Rico2020,e-collpase-Garbe-2017, Irish2014,Irish2017,PRX-Xie-Anistropy,Batchelor2015,XieQ-2017JPA,Hwang2015PRL,Bera2014Polaron,Ying2015,LiuM2017PRL, Ying-2018-arxiv,Ying-2021-AQT,CongLei2017,CongLei2019,Ying2020-nonlinear-bias,Liu2021AQT,ChenQH2012,e-collpase-Duan-2016, ZhangYY2016,ZhengHang2017,PengJie2019,Liu2015,Ashhab2013,ChenGang2012,FengMang2013,Eckle-2017JPA,Casanova2018npj, HiddenSymMangazeev2021,HiddenSymLi2021,HiddenSymBustos2021,Garbe2020,
Garbe2021-Metrology,Boite2016-Photon-Blockade,Ridolfo2012-Photon-Blockade,Irish-class-quan-corresp,Li2020conical,Ying-gapped-top,Braak2019Symmetry,Ma2020Nonlinear,Ilias2022-Metrology}
Without mentioning the ubiquitous role of light-matter interaction and its
broad relevance to quantum optics, quantum information and quantum
computation,\cite%
{Diaz2019RevModPhy,Romero2012,Stassi2020QuComput,Stassi2018,Macri2018}
quantum metrology,\cite{Garbe2020,Garbe2021-Metrology,Ilias2022-Metrology,Ying2022-Metrology} condensed matter,\cite%
{Kockum2019NRP,Ying-2021-AQT,Ying-gapped-top} and relativistic systems,\cite%
{Bermudez2007} the explosively-growing investigations have yielded abundant
findings in the QRM and its extensions, such as hidden symmetry,\cite%
{Braak2019Symmetry,HiddenSymMangazeev2021,HiddenSymLi2021,HiddenSymBustos2021}
various patterns of symmetry breaking,\cite%
{Ying2020-nonlinear-bias,Ying-2018-arxiv,Ying-2021-AQT} few-body quantum
phase transitions,\cite%
{Liu2021AQT,Ashhab2013,Ying2015,Hwang2015PRL,Ying2020-nonlinear-bias,Ying-2021-AQT,LiuM2017PRL,Hwang2016PRL,Ying-gapped-top,Ying-2018-arxiv}
multicriticalities and multiple points,\cite%
{Ying2020-nonlinear-bias,Ying-2021-AQT,Ying-gapped-top} universality
classification,\cite{Hwang2015PRL,LiuM2017PRL,Irish2017,Ying-2021-AQT}
spectral collapse,\cite%
{Felicetti2015-TwoPhotonProcess,e-collpase-Garbe-2017,e-collpase-Duan-2016,CongLei2019,Rico2020}
photon blockade effect,\cite%
{Boite2016-Photon-Blockade,Ridolfo2012-Photon-Blockade} spectral conical
intersections,\cite{Li2020conical} classical-quantum correspondence,\cite%
{Irish-class-quan-corresp} single-qubit conventional and unconventional
topological phase transitions,\cite{Ying-2021-AQT,Ying-gapped-top} and so
forth.

An intriguing phenomenon most relevant in coupling enhancement may be quantum
phase transition (QPT).\cite%
{Liu2021AQT,Ashhab2013,Ying2015,Hwang2015PRL,Ying2020-nonlinear-bias,Ying-2021-AQT,LiuM2017PRL,Hwang2016PRL}
Generally speaking, QPTs are transitions of ground states (GSs) induced by a
variation of some non-thermal parameter.\cite{Sachdev-QPT} In contrast to thermal fluctuations
in classical phase transitions, QPTs are regarded to be driven by quantum
fluctuations and traditionally lie in the thermodynamic limit in condensed
matter. Interestingly, the QRM as a few-body system also exhibits a QPT \cite%
{Ashhab2013,Ying2015,Hwang2015PRL} in the low-frequency limit, i.e., $\omega
/\Omega \rightarrow 0$ where $\omega $ is the bosonic frequency and $\Omega $
is the atomic level splitting or tunneling strength. It was also suggested
that whether the transition should be termed quantum or not is a matter of
taste by taking the negligible quantum fluctuations in the photon vacuum
state into account.\cite{Irish2017} Nevertheless, when critical universality
is a character often born with QPTs as in the condensed matter, it has been
shown that the anisotropic QRM manifests a universal scaling relation in the
critical exponent that can be really bridged to the thermodynamic limit.\cite%
{LiuM2017PRL}

Opposite to universality is diversity which represents the quality to be
diverse or different. With the opposite qualities universality and diversity
are apparently antagonists. Unexpectedly, the universality scenarios in the
anisotropic QRM demonstrate that they can turn to support each other.
Indeed, the afore-mentioned critical universality of scaling relation needs
the condition of low frequency limit, while at finite frequencies the
universal scaling relation breaks down and the system properties are
dominated by diversity. However, amidst the diversity a new universality
classification can be found from the topological structure of the
GS wave function.\cite{Ying-2021-AQT} In fact, such
universality-diversity-universality scenarios involve two different kinds of
universalities: one is critical universality, while the other is topological
universality. Note that such scenarios occur in the anisotropic QRM which is
linear in the light-matter interaction, one may wonder whether the
universalities are simply a special case or hold more generally, e.g., in a
nonlinear coupling.

To get more robust universalities we consider the QRM with both anisotropy
and the nonlinear Stark coupling in the present work. We consider both the
low-frequency limit and the finite-frequency case. In the low-frequency
limit we analytically obtain the phase boundaries of QPTs and extract
different levels of scaling relations which are valid respectively in
various anisotropic couplings or for both anisotropic and nonlinear Stark
couplings, locally around transitions or globally for all critical regimes.
At finite frequencies, indeed the critical universality collapses and
diversity dominates, while topological phase transitions (TPTs) emerge. Both
conventional and unconventional TPTs respectively with and without gap
closing are present. Their different sensitivities in response to the
nonlinear Stark coupling enable the forming of topological quadruple points,
while it never occurs in linear interaction. A further analysis by composite
phase diagrams with hexaple points reveals that the topological quadruple
points are actually spin-invariant points.

The paper is organized as follows. Section \ref{Sect-Model} introduces the
anisotropic QRM with nonlinear Stark coupling and addresses the symmetry in
quadrature representation. In Section \ref{Sect-QPTs-low-w} methods are
introduced to obtain analytic boundaries of QPTs in the low-frequency limit.
Different levels of scaling relations are extracted. Section \ref%
{Sect-Diversity} shows the breakdown of the critical universality and
arising of diversity at finite frequencies. Section \ref%
{Sect-Top-Classification} presents topological classifications at
finite frequencies, with findings of topological quadruple points, composite
hexaple points and invariant points. Section \ref{Sect-Mechanisms} is
devoted to mechanism clarifications. Conclusions and discussions are finally
given in Section \ref{Sect-Conclusions}.

\section{Model and Symmetry}

\label{Sect-Model}

The standard QRM has a linear and isotropic interaction, while in
experimental setups extended versions of QRM are often applied. Indeed,
coupling anisotropy plays an important role in ultrastrong couplings \cite%
{PRX-Xie-Anistropy,Forn-Diaz2010} and is highly tunable.\cite{Yimin2018} On
the other hand, a so-called Stark nonlinear coupling can be added and
realized with adjustable amplitude and sign.\cite%
{Eckle-2017JPA,Stark-Grimsmo2013,Stark-Grimsmo2014,Stark-Cong2020} We
consider the QRM with both anisotropy and nonlinear Stark coupling as
described by the following Hamiltonian
\begin{eqnarray}
H &=&\omega a^{\dagger }a+\frac{\Omega }{2}\sigma _{x}+\chi \omega \hat{n}%
\sigma _{x}+H_{g},  \label{Ha} \\
H_{g} &=&g\left[ \left( \widetilde{\sigma }_{-}a^{\dagger }+\widetilde{%
\sigma }_{+}a\right) +\lambda \left( \widetilde{\sigma }_{+}a^{\dagger }+%
\widetilde{\sigma }_{-}a\right) \right] .
\end{eqnarray}%
Here $\omega $ is the frequency of a bosonic mode created (annihilated) by $%
a^{\dagger }$ $(a)$, while $\Omega $ is atomic level splitting in cavity
systems or tunneling strength in superconducting circuit systems with the
qubit (spin) represented by the Pauli matrix $\sigma _{x,y,z}$. The linear coupling
strength is controlled by $g$. The anisotropy $\lambda $ tunes the ratio of the rotating-wave terms and the
counter-rotating terms, with $\lambda
=1$ and $\lambda =0$ retrieving the QRM \cite{rabi1936} and the
Jaynes-Cummings model (JCM) \cite{JC-model} respectively. Note we have
adopted the spin notation as in ref.\cite{Irish2014}, in which $\sigma
_{z}=\pm $ labels two flux states in flux-qubit circuit systems.\cite%
{flux-qubit-Mooij-1999} In such a spin notation, the spin raising and
lowering operators on $\sigma _{x}$ basis are expressed by $\widetilde{%
\sigma }^{\pm }=(\sigma _{z}\mp i\sigma _{y})/2$, while one can recover the
conventional form by a spin rotation \{$\sigma _{x},\sigma _{y},\sigma _{z}$%
\} $\rightarrow $ \{$\sigma _{z},-\sigma _{y},\sigma _{x}$\} around the axis
$\vec{x}+\vec{z}$. \ The $\chi $ term denotes the nonlinear Stark coupling
with a limitation $\left\vert \chi \right\vert \leqslant 1$ beyond which the
system energy would be negatively unbound thus unphysical.

It should be noted that neither the anisotropy nor the nonlinear Stark
coupling breaks the parity so that the model preserves the parity symmetry,
with $H$ commuting with the parity operator $\hat{P}=\sigma
_{x}(-1)^{a^{\dagger }a}$. The parity symmetry is relevant for
symmetry-protected TPTs \cite{Ying-2021-AQT,Ying-gapped-top} as in condensed matter,\cite{Topo-Wen,TopCriterion,Top-Guan,TopNori} while there is a hidden symmetry breaking of spin
reversion or space inversion for the symmetry-breaking QPT in the GS.\cite{Ying-2021-AQT}

Changing to the quadrature representation by $a^{\dagger }=(\hat{x}-i\hat{p}%
)/\sqrt{2},$ $a=(\hat{x}+i\hat{p})/\sqrt{2}$ with momentum $\hat{p}=-i\frac{%
\partial }{\partial x}$ will facilitate our analysis in the effective position
space
\begin{eqnarray}
H &=&\frac{\omega }{2}\hat{p}^{2}+v_{\sigma _{z}}\left( x\right)
+H_{+}\sigma ^{+}+H_{-}\sigma ^{-},  \label{Hx} \\
H_{\pm } &=&\frac{\left( \Omega -\chi \omega \right) }{2}\mp g_{y}i\sqrt{2}%
\hat{p}+\frac{\chi \omega }{2}\left( \hat{x}^{2}+\hat{p}^{2}\right) .
\end{eqnarray}%
It should be noted here that, differently from $\widetilde{\sigma }^{\pm }$,
the spin raising and lowering in $\sigma _{x}=\sigma ^{+}+\sigma ^{-}$, $%
\sigma _{y}=-i(\sigma _{+}-\sigma _{-})$ are now on $\sigma _{z}=\pm $
basis. We have defined $g_{y,z}^{\prime }=\sqrt{2}g_{y,z}/\omega $ for $%
g_{y}=\frac{\left( 1-\lambda \right) }{2}g$ and $g_{z}=\frac{\left(
1+\lambda \right) }{2}g,$ thus $g_{z}^{\prime }$ is effectively the
amplitude of spin-dependent displacement in harmonic potentials $v_{\sigma
_{z}}\left( x\right) =\omega \left( x+g_{z}^{\prime }\sigma _{z}\right)
^{2}/2+\varepsilon _{0}^{z}$ where $\varepsilon _{0}^{z}=-\frac{1}{2}%
[g_{z}^{\prime 2}+1]\omega $. In such a representation the $\Omega $ term
effectively plays the role of spin flipping in the spin $\sigma _{z}$ space
and the role of tunneling in the effective position space.\cite%
{Ying2015,Irish2014} The $g_{y}$ term takes the form $\sqrt{2}g_{y}\hat{p}%
\sigma _{y}$ as the Rashba spin-orbit coupling in nanowires \cite%
{Nagasawa2013Rings,Ying2016Ellipse,Ying2017EllipseSC} or the equal-weight mixture \cite%
{LinRashbaBECExp2011,LinRashbaBECExp2013Review} of the linear
Dresselhaus ($\hat{p}_{x}\sigma _{y}+\hat{p}%
_{y}\sigma _{x}$) and Rashba ($\hat{p}_{x}\sigma _{y}-\hat{%
p}_{y}\sigma _{x}$) spin-orbit couplings in condensed matter\cite{Dresselhaus1955,Rashba1984} and cold atomic gases.\cite{LinRashbaBECExp2011,LinRashbaBECExp2013Review,Li2012PRL}

The Hamiltonian can be rewritten in $x$-$p$ dual forms
\begin{eqnarray}
H_{x} &=&\frac{\omega }{2}[(-i\frac{\partial }{\partial x}+g_{y}^{\prime
}\sigma _{y})^{2}+\left( x+g_{z}^{\prime }\sigma _{z}\right) ^{2}]  \nonumber
\\
&&+[\frac{\left( \Omega -\chi \omega \right) }{2}+\frac{\chi \omega }{2}%
\left( \hat{x}^{2}+\hat{p}^{2}\right) ]\sigma _{x}+\varepsilon _{0},
\label{H2-x} \\
H_{p} &=&\frac{\omega }{2}[(-i\frac{\partial }{\partial p}-g_{z}^{\prime
}\sigma _{z})^{2}+\left( p+g_{y}^{\prime }\sigma _{y}\right) ^{2}]  \nonumber
\\
&&+[\frac{\left( \Omega -\chi \omega \right) }{2}+\frac{\chi \omega }{2}%
\left( \hat{x}^{2}+\hat{p}^{2}\right) ]\sigma _{x}+\varepsilon _{0},
\label{H2-p}
\end{eqnarray}%
where $\varepsilon _{0}=-\omega (1+g_{z}^{\prime 2}+g_{y}^{\prime 2})/2$ and
$\hat{x}=i\frac{\partial }{\partial p}$. From from $H_{x}$ and $H_{p}$ one
sees that $\lambda >0$ and $\lambda <0$ regimes are symmetric under the spin
rotation and transform to momentum space $\left\{ \sigma _{x},\sigma
_{y},\sigma _{z}\right\} \rightarrow \left\{ \sigma _{x},-\sigma _{z},\sigma
_{y}\right\} $, $ x\rightarrow p $, $ \lambda \rightarrow -\lambda $.\cite{Ying-2021-AQT}
Phase transitions in the GS would involve a linear coupling of
order as the critical point in the absence of the nonlinear Stark coupling $%
g_{c}^{\lambda }=\frac{2}{1+\left\vert \lambda \right\vert }g_{\mathrm{s}}$,%
\cite{LiuM2017PRL,Ying-2021-AQT} with $g_{\mathrm{s}}=\sqrt{\omega \Omega }%
/2,$ in the low frequency limit $\omega /\Omega \rightarrow 0$ or that of a
TPT, $g_{\mathrm{T1}}^{\lambda }=\frac{2}{\sqrt{1-\lambda ^{2}}}g_{\mathrm{s}%
},$\cite{Ying-2021-AQT} at finite frequencies. At low frequencies, the
contribution of $v_{\sigma _{z}}\left( x\right) $ and $\hat{x}^{2}$ terms
are of leading order \cite{Ying2020-nonlinear-bias} $\Omega $\ while the $%
g_{y}$ term and $\hat{p}^{2}$ term are of subdominant orders $\left( \omega
\Omega \right) ^{1/2}$ and $\omega ^{1}$. Thus, the $\lambda >0$ regime is $x
$-type in the sense $\langle \hat{x}^{2}\rangle $ is more dominant than $%
\langle \hat{p}^{2}\rangle $, with the main characters more conveniently
described by $H_{x}$. At finite frequencies, the $g_{y}$ term has some
self-cancelation effect due to oscillation as seen later on, while larger
amplitudes of $g_{z}^{\prime }$ than $g_{y}^{\prime }$ still favor an $x$-type
state in $\lambda >0$ regime. Hereafter, unless specially mentioned, we
shall focus on $\lambda >0$ regime while one has similar results with a $p$%
-type state by $H_{p}$ in the momentum space for $\lambda <0$ regime.

\section{QPTs and Scaling Relations in Low-Frequency Limit}

\label{Sect-QPTs-low-w}

\begin{figure*}[tbph]
\centering
\includegraphics[width=1.8%
\columnwidth]{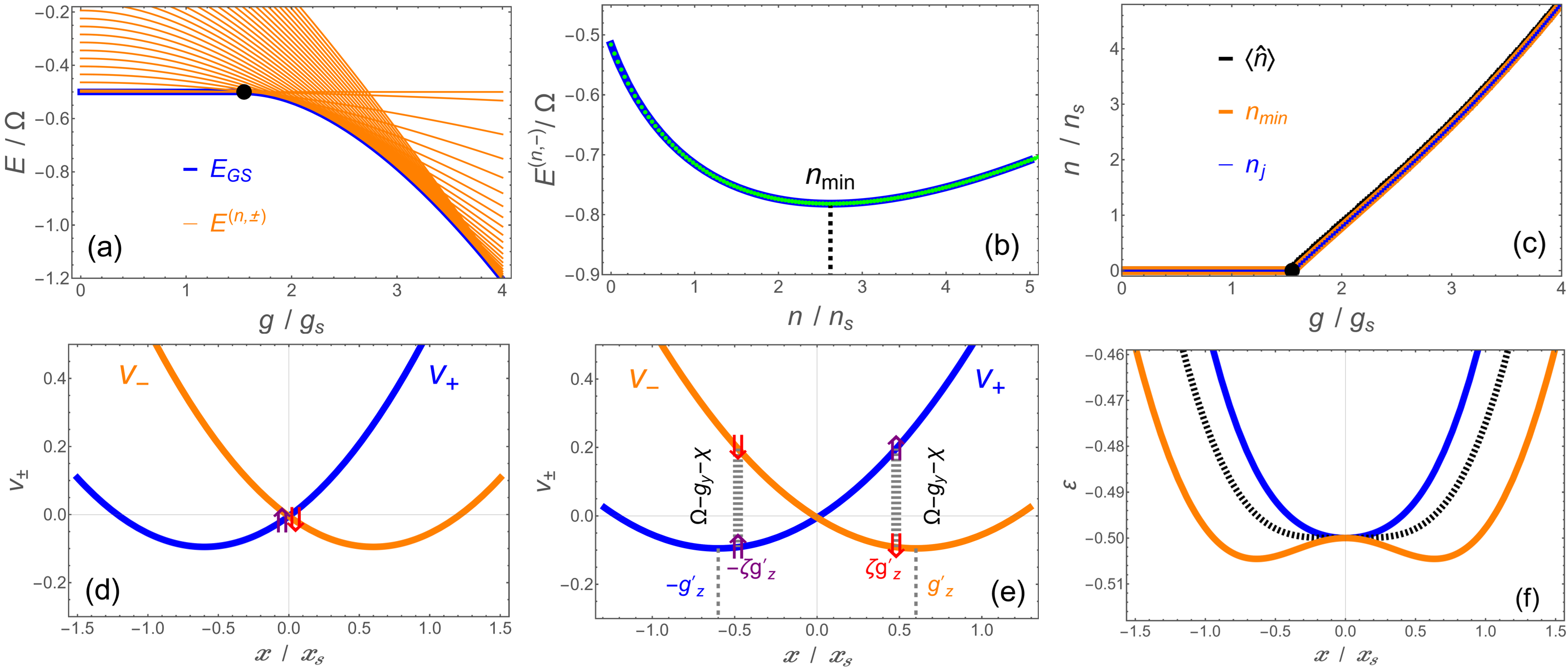}
\caption{\textit{Exact solution at $\lambda =0$ (a-c) and
variational method for $\lambda \neq 0$ (d-f) in low-frequency limit with nonlinear Stark coupling.%
} Here $\omega =0.01\Omega $, $\chi =0.4$ for (a-b) and $%
\lambda =0.5$ for (d-f). (a) Analytic energy spectrum versus $g$ for
excited states $E^{n,-}$ (orange thin lines) and ground state (GS) $E_{GS}$
(blue thick line). (b) $E^{n,-}$ as a continuous (blus solid) or discrete
(green dots) function of $n$ at $g=3g_{\mathrm{s}}$, with optimized number $%
n_{min}$ marked by the vertical dashed line. (c) Discrete quantum number $n_{j}$ in $%
E^{(n_{j},-)}$ (blue), optimized number $n_{min}$ (orange) and the
expectation of photon number $\langle \hat{n}\rangle $ (black) at $g=3g_{%
\mathrm{s}}$ for the GS. The dots in (a,c) mark the transition points $%
g_{c}^{\lambda ,\chi }$ in the low-frequency limit. (d) Spin
configuration before transition at $\chi =0.54$, the arrows mark
the positions in the effective potential for spin-up (blue, $v_{+}$)
spin-down (orange, $v_{-}$ components. (e) Spin configuration after
transition in $v_{\pm }$ at $\chi =0.74$ with variational
displacements renormalized by $\zeta $ from the potential-bottom
positions $\pm g_{z}^{\prime }$ (dashed). (f) Variational energy $%
\varepsilon $ versus the displacement $x$ at $g=0.8g_{\mathrm{s}}$ and $%
\lambda =0.5$, for $\chi =0.54$ (blue, upper), $0.64$
(dashed, middle), $0.74$ (orange, lower) and the critical point is $%
\chi _{c}=0.64$. }
\label{fig-Vari-eSC}
\end{figure*}

We shall first study the low-frequency limit to extract GS phase diagrams and critical scaling relations.
We figure out the full phase diagrams numerically by the exact diagonalization,\cite{Ying2020-nonlinear-bias}
while to obtain analytic phase boundaries and find different scaling relations we need some analytic methods. For the latter
purpose, we fall back on a semiclassical variational method for $\lambda \neq 0$ and the exact solution at $\lambda =0$, as described in this
section. We will get different levels of scaling relations and eliminate a singular behavior at $\lambda =0$. The obtained analytic phase boundaries will also provide a reference to fix the invariant points at finite frequencies in next section.

\subsection{Explicit Solution and Energy at $\lambda =0$}

\subsubsection{General Solution at Any Frequencies}

The explicit exact solution is available for the JCM at $\lambda =0$ in linear coupling,\cite{JC-model,JC-Larson2021}
while here we shall address in the presence of the nonlinear
Stark coupling. Setting $\lambda =0$ drops the counter-rotating terms in the
Hamiltonian (\ref{Ha}) so that the eigenstates only involve at most two
bases in the following form%
\begin{eqnarray}
\psi _{n}^{\left( \pm \right) } &=&\left( C_{n\Uparrow }^{\left( \pm \right)
}\left\vert n,\Uparrow \right\rangle _{\sigma _{x}}+C_{n\Downarrow }^{\left(
\pm \right) }\left\vert n+1,\Downarrow \right\rangle _{\sigma _{x}}\right)
/N,  \label{WaveF-JC-n} \\
\psi _{0} &=&\left\vert 0,\Downarrow \right\rangle _{\sigma _{x}}
\end{eqnarray}%
where $\Uparrow ,\Downarrow $ are spins states of $\sigma _{x}$ as labeled
by the subscript of the basis. The coefficients on the basis are explicitly
given by%
\begin{eqnarray}
C_{n\Uparrow }^{\left( \pm \right) } &=&e_{-}\pm \sqrt{e_{-}^{2}+\left(
n+1\right) g^{2}}, \\
C_{n\Downarrow }^{\left( \pm \right) } &=&g\sqrt{\left( n+1\right) },
\end{eqnarray}%
where $e_{+}=\left( n+\frac{1-\chi }{2}\right) \omega $ and $e_{-}=\frac{1}{2%
}\left( \Omega -\omega \right) +\left( n+\frac{1}{2}\right) \chi \omega $
and $N=C_{n\Uparrow }^{\left( \pm \right) 2}+C_{n\Downarrow }^{\left( \pm
\right) 2}$ is the normalization factor.\ Corresponding to the above states
one can get the eigenenergies%
\begin{eqnarray}
E_{\mathrm{JC-Stark}}^{\left( n,\pm \right) } &=&e_{+}\pm \sqrt{%
e_{-}^{2}+\left( n+1\right) g^{2}}, \\
E_{\mathrm{JC-Stark}}^{0} &=&-\frac{\Omega }{2},
\end{eqnarray}%
respectively. Note the energy in branch $E_{\mathrm{JC-Stark}}^{\left(
n,+\right) }$ is higher than $E_{\mathrm{JC-Stark}}^{\left( n,-\right) }$,
thus the GS lies in the competition among the corresponding
states in branch $\psi _{n}^{\left( -\right) }$ as well as $\psi _{0}$, as
shown in Fig. \ref{fig-Vari-eSC} where the orange thin lines are $E_{\mathrm{%
JC-Stark}}^{\left( n,+\right) }$ (only integer-$n/5$ plotted), with $E_{%
\mathrm{JC-Stark}}^{0}$ being the horizontal one, while the final GS is
indicated by the blue line.

\subsubsection{Photon Number in Low-Frequency Limit}

In the low-frequency limit $\omega /\Omega \rightarrow 0$, the spacing of
the quantum number $n$ becomes small relatively to the characteristic number
$n_{\mathrm{s}}=x_{\mathrm{s}}^{2}/2=\Omega /(4\omega ),$ where $x_{\mathrm{s%
}}=\sqrt{2}g_{\mathrm{s}}/\omega =\sqrt{\Omega /(2\omega )}$, which is the
order of photon number induced by a coherent state in the displaced
potential. In such a situation we can approximately regard the energy $E_{%
\mathrm{JC-Stark}}^{\left( n,-\right) }$ as a continuous function of $n$. As
illustrated in Fig. \ref{fig-Vari-eSC}b, the minimization of $E_{\mathrm{%
JC-Stark}}^{\left( n,-\right) }$ with respect to $n$ gives an optimal
quantum number%
\begin{equation}
n_{\min }=\frac{1-\chi }{2\chi }-\frac{\left( \overline{g}_{\mathrm{s}%
}^{2}+4\chi \right) \Omega }{8\chi ^{2}\omega }+\frac{\overline{g}_{\mathrm{s%
}}\Omega }{8\chi ^{2}\omega }\sqrt{\frac{\overline{g}_{\mathrm{s}}^{2}+8\chi
_{1}}{1-\chi ^{2}}},  \label{nMin}
\end{equation}%
with $\chi _{1}=\chi \lbrack 1-(1+\chi )\frac{\omega }{\Omega }]$.

The integer number $n_{j}$ nearest to $n_{\min }$ should be the discrete
quantum number of the GS after the transition which can also obtained by
level crossing $E_{\mathrm{JC-Stark}}^{\left( n,-\right) }=E_{\mathrm{%
JC-Stark}}^{\left( n+1,-\right) }$ at
\begin{eqnarray}
\overline{g}_{\mathrm{s}} &=&\sqrt{-\chi +C_{1}\frac{\omega }{\Omega }+\sqrt{%
1-2\left( 1+\chi \right) \frac{\omega }{\Omega }+C_{2}\frac{\omega ^{2}}{%
\Omega ^{2}}}}, \\
n_{j} &=&\frac{1-2\chi }{2\chi }-\frac{\left( \overline{g}_{\mathrm{s}%
}^{2}+4\chi \right) \Omega }{8\chi ^{2}\omega }+\frac{\overline{g}_{\mathrm{s%
}}\Omega }{8\chi ^{2}\omega }\sqrt{\frac{\overline{g}_{\mathrm{s}}^{2}+8\chi
_{1}+d}{1-\chi ^{2}}},
\end{eqnarray}%
where $d=16\chi ^{2}(1-\chi ^{2})\omega ^{2}/(\overline{g}_{\mathrm{s}%
}^{2}\Omega ^{2})$, $C_{1}=(1+\chi )[3-2\chi +2\left( 1-\chi \right) n]/2,$ $%
C_{2}=(1+\chi )[9-7\chi +4(1-\chi )n(n+3)].$

In the low-frequency limit, both $n_{\min }$ and $n_{j}$ approach to%
\begin{equation}
\frac{n_{\min }}{n_{\mathrm{s}}}=\frac{n_{j}}{n_{\mathrm{s}}}=-\frac{%
\overline{g}_{\mathrm{s}}^{2}+4\chi }{4\chi ^{2}}+\frac{\overline{g}_{%
\mathrm{s}}}{4\chi ^{2}}\sqrt{\frac{\overline{g}_{\mathrm{s}}^{2}+8\chi }{%
1-\chi ^{2}}}.  \label{nMin-low-w}
\end{equation}%
Figure \ref{fig-Vari-eSC}c compares $n_{\min }$ (orange)$,n_{j}$ (blue) with
the expectation of photon number $\langle \widehat{n}\rangle =[nC_{n\Uparrow
}^{\left( \pm \right) 2}+(n+1)C_{n\Downarrow }^{\left( \pm \right) 2}]/N$
(black) at a frequency $\omega =0.01\Omega $, they all coincide. The result (%
\ref{nMin-low-w}) will be later on used for discussion on the singular point
in scaling relation.

\subsection{Semiclassical Variational Method for $\lambda >0$}

\label{Sect-E-semiclassic}

Now let us discuss the QPTs in $\lambda >0$ regime, while one gets the
same results for $\lambda <0$ regime in the momentum space. Note the
critical coupling is of order $g_{\mathrm{s}}$ which yields a potential
energy $v_{\sigma _{z}}$ of order $\Omega $. In the low frequency limit, the
kinetic energy is of order $\omega $, thus being relatively negligible. The
Rashba spin-orbit coupling term has a strength $g_{y}$ which is of order $\omega
^{1/2}$ at critical couplings, also being negligible. Thus, in the leading
order, we can reduce the model to a semiclassical Hamiltonian for the GS
which has a zero momentum as the GS of a classical particle
while the quantum part is kept in spin space:\cite{Ying2020-nonlinear-bias}
\begin{equation}
H_{x}\rightarrow H_{\mathrm{SC}}^{x}=\frac{\omega }{2}\left( x+g_{z}^{\prime
}\sigma _{z}\right) ^{2}+\varepsilon _{\mathrm{SC}}^{z}+\left( \frac{\Omega
}{2}+\frac{\chi \omega }{2}x^{2}\right) \sigma _{x},  \label{H-SemiClassical}
\end{equation}%
where $\varepsilon _{\mathrm{SC}}^{z}=-%
\frac{1}{2}g_{z}^{\prime 2}\omega $. For the semiclassical approximation we have drop the zero-point energy
$\frac{\omega }{2}$ in $\hat{n}\omega =\frac{\omega }{2}\left( \hat{x}^{2}+%
\hat{p}^{2}\right) -\frac{\omega }{2}$ and the Stark term. From an alternative
angle $\frac{\omega }{2}$ and $\frac{\chi \omega }{2}$ also can be drop due
to negligible order in the low-frequency limit. One gets similar reduced
Hamiltonian $H_{\mathrm{SC}}^{p}$ of $H_{p}$ for $\lambda <0$ regime with
the position $x$ replaced by the momentum $p$ and $g_{z}^{\prime }$ changed
to be $g_{y}^{\prime }$. We see that at infinity $x\rightarrow \infty $ the
energy would be dominated by
\begin{equation}
H_{\mathrm{SC}}\rightarrow \frac{\omega }{2}\left( 1+\chi \sigma _{x}\right)
x^{2}+\varepsilon _{\mathrm{SC}}^{z}\
\end{equation}%
which is negatively unbound for $\left\vert \chi \right\vert >1,$ thus this
regime is unstable and unphysical as mentioned in Section \ref{Sect-Model}. Hereafter we shall focus on the physical
regime $\left\vert \chi \right\vert \leqslant 1$.

To obtain the explicit energy one can rewrite $H_{\mathrm{SC}}$ in a matrix
form
\begin{equation}
H_{\mathrm{SC}}=\left(
\begin{array}{cc}
h_{\uparrow \uparrow } & h_{\uparrow \downarrow } \\
h_{\downarrow \uparrow } & h_{\downarrow \downarrow }%
\end{array}%
\right) ,
\end{equation}%
where
\begin{eqnarray}
h_{\uparrow \uparrow } &=&\frac{\omega }{2}\left( x+g_{z}^{\prime }\right)
^{2}+\varepsilon _{\mathrm{SC}}^{z},  \nonumber \\
h_{\uparrow \uparrow } &=&\frac{\omega }{2}\left( x-g_{z}^{\prime }\right)
^{2}+\varepsilon _{\mathrm{SC}}^{z}, \\
h_{\uparrow \downarrow } &=&h_{\downarrow \uparrow }=\frac{\Omega}{2}+\frac{\chi \omega }{2}x^{2}.  \nonumber
\end{eqnarray}%
Diagonalization of $H_{\mathrm{SC}}$ gives two energies with the lower one being
\begin{equation}
\varepsilon =\frac{\omega }{2}\left( x^{2}+g_{z}^{\prime 2}\right) -\frac{%
\omega }{2}\sqrt{4g_{z}^{\prime 2}x^{2}+\left( x^{2}\chi +\frac{\Omega }{%
\omega }\right) ^{2}}+\varepsilon _{\mathrm{SC}}^{z}  \label{E-SC}
\end{equation}%
which is still variational as the spatial position $x$ has not yet been
optimized. Minimization with respect to $x$
\begin{equation}
\frac{\partial \varepsilon }{\partial x}=0
\end{equation}%
leads to two solutions for the most favorable position
\begin{eqnarray}
x_{m}^{B} &=&0, \\
\left\vert x_{m}^{A}\right\vert  &=&\sqrt{\frac{2g_{z}^{\prime }}{\chi ^{2}}%
\sqrt{\frac{\left( g_{z}^{\prime 2}+\chi \frac{\Omega }{\omega }\right) }{%
\left( 1-\chi ^{2}\right) }}-\frac{\left( 2g_{z}^{\prime 2}+\chi \frac{%
\Omega }{\omega }\right) }{\chi ^{2}}}.
\end{eqnarray}%
After substituting $\left\vert x_{m}\right\vert $ into (\ref{E-SC}) we
arrive at the explicit final energies%
\begin{eqnarray}
E_{\mathrm{SC}}^{B} &=&-\frac{\Omega }{2}, \\
E_{\mathrm{SC}}^{A} &=&\varepsilon _{\mathrm{SC}}^{z}-\frac{g_{z}^{\prime
2}\left( 2-\chi ^{2}\right) \omega +\chi \Omega }{2\chi ^{2}}  \nonumber \\
&&+\frac{g_{z}^{\prime }\left( 1-\chi ^{2}\right) \omega }{\chi ^{2}}\sqrt{%
\frac{g_{z}^{\prime 2}+\chi \frac{\Omega }{\omega }}{1-\chi ^{2}}}
\label{Ea-SC}
\end{eqnarray}%
which actually are the energies before and after the phase transition,
respectively, as discussed in the following.

\subsection{Phase Diagrams and Critical Scalings in Low-Frequency limit}

\label{Sect-phase-diagram-low-w}

\subsubsection{Phase Transition and Critical Boundary}

Actually at a small coupling strength $\left\vert x_{m}^{A}\right\vert $ is
imaginary so the only physical solution is $x_{m}^{B}$. Indeed, in such a
situation $\varepsilon $ has a single minimum which is located at the
origin, as illustrated by the blue (upper) line in Figure \ref{fig-Vari-eSC}f.
The variational energy remains in the mono-minimum profile till a second-order phase
transition is triggered at a critical point
\begin{eqnarray}
g_{c}^{\lambda ,\chi } &=&\frac{2\sqrt{\left( 1-\chi \right) }}{1+\left\vert
\lambda \right\vert }g_{\mathrm{s}}=\sqrt{\left( 1-\chi \right) }%
g_{c}^{\lambda },  \label{gC-SC} \\
\left\vert \lambda _{c}^{\lambda ,\chi }\right\vert  &=&\frac{2\sqrt{\left(
1-\chi \right) }}{g/g_{\mathrm{s}}}-1,  \label{lambdaC-SC} \\
\left\vert \chi _{c}^{\lambda ,\chi }\right\vert  &=&1-\left( 1+\left\vert
\lambda \right\vert \right) ^{2}\frac{g^{2}}{4g_{\mathrm{s}}^{2}},
\label{ChiC-SC}
\end{eqnarray}%
after which a double-minimum structure shows up. The black dotted line in
Figure \ref{fig-Vari-eSC}f shows the case right at the critical point, with a
flat bottom at the origin. After the critical point, as demonstrated by the
orange (lower) line, two degenerate minima appear
at $\pm \left\vert x_{m}^{A}\right\vert $ while $x_{m}^{B}$ becomes an
unstable local maximum at the origin. Note the expressions for the critical
point are general for both $\chi $ and $\lambda $, setting $\chi =0$
recovers the boundary for the anisotropic QRM~\cite{LiuM2017PRL,Ying-2021-AQT}
\begin{equation}
g_{c}^{\lambda }=\frac{2}{1+\left\vert \lambda \right\vert }g_{\mathrm{s}}
\end{equation}%
in the absence of the Stark coupling.
Setting $\lambda =0$, the general critical boundary $g_{c}^{\lambda ,\chi }$
reduces to the Stark-JC critical point%
\begin{equation}
g_{c}^{\mathrm{JC-Stark}}=2g_{\mathrm{s}}\sqrt{\left( 1-\chi \right) }
\label{gc-JC-Stark}
\end{equation}%
which is also exactly obtained by the level crossing $E_{\mathrm{JC-Stark}%
}^{\left( 0,-\right) }=E_{\mathrm{JC-Stark}}^{0}$.

\begin{figure*}[tbph]
\centering
\includegraphics[width=2.0%
\columnwidth]{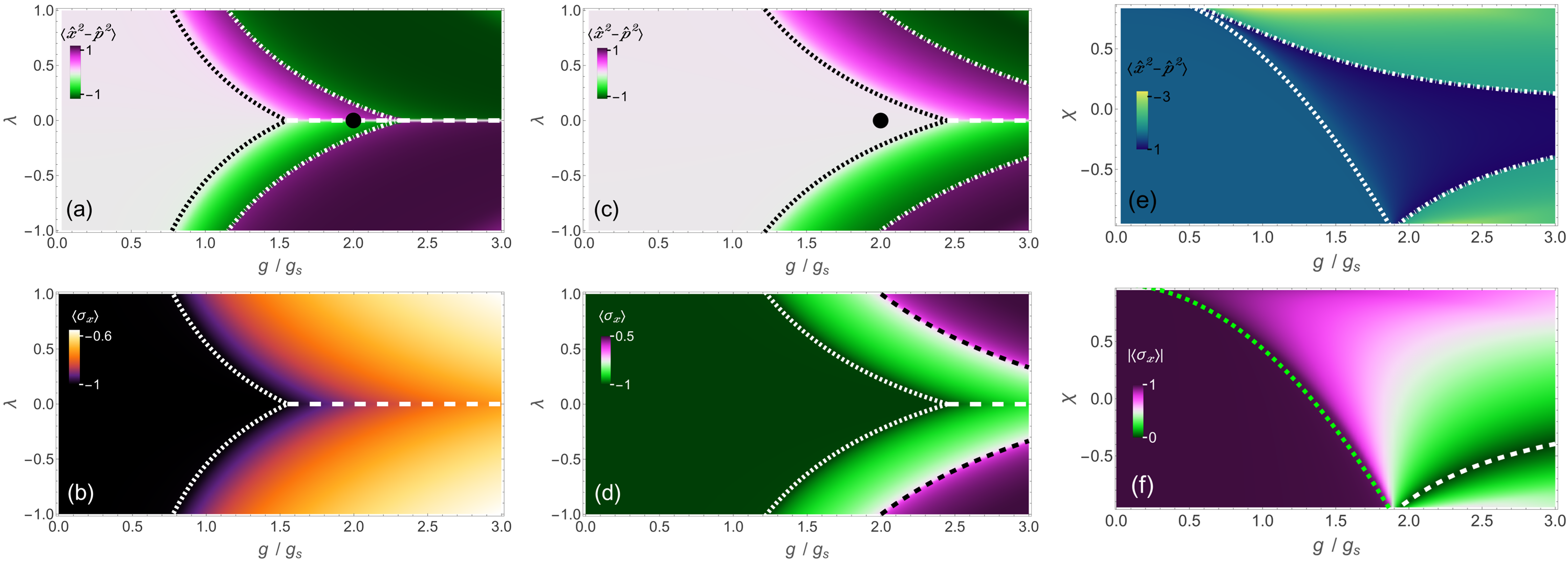}
\caption{\textit{GS phase diagrams in the low-frequency limit.}
Density plots in the $g$-$\lambda $ plane and the $g$-$\chi $
plane for $\langle \hat{x}^{2}\rangle -\langle \hat{p}^{2}\rangle $ (a,c,e) and
the spin expectation $\langle \sigma _{x}\rangle $ (b,d,f) with a
fixed $\chi =0.4$ (a,b), $\chi =-0.5$ (c,d), and $%
\lambda =0.5$ (e,f).  Here $\omega =0.01\Omega $. To visualize the numerical boundaries better,
$\langle \hat{x}^{2}\rangle -\langle \hat{p}^{2}\rangle $ is scaled by
$x_{z,y}^2$ and timed with
$\text{sign}[1-|\langle \hat{x}^{2}\rangle -\langle \hat{p}^{2}\rangle| /x_{z,y}^2]$, while in (d) the amplitude is
amplified by $|\langle \sigma _{x}\rangle|^{1/3}$. Apart from the JC line (horizontal long-dashed), the dotted, dashed,
and dot-dashed curves are analytic boundaries $g_{c}^{\lambda ,%
\chi }$ (\ref{gC-SC}), $g_{c}^{\sigma _x}$ (
\ref{gc-SpinX}), and $g_{c}^{\zeta i}$ (\ref{gC-adiabtic-1},%
\ref{gC-adiabtic-2}), respectively. }
\label{fig-diagrams-low-w}
\end{figure*}

In \textbf{Figure} \ref{fig-diagrams-low-w} we show the phase diagrams of the expectation
$\langle \hat{x}^{2}\rangle -\langle \hat{p}^{2}\rangle $ (a,c,e) and the spin expectation
$\langle \sigma _{x}\rangle $ (b,d,f) at a low frequency $\omega =0.01\Omega $.
The dotted curves denote the analytic critical boundary (\ref{gC-SC})-(\ref%
{ChiC-SC}) which agrees well with the second-order-like transition in numerics.
Figure \ref{fig-diagrams-low-w}a,c indicates that
$\langle \hat{x}^{2}\rangle -\langle \hat{p}^{2}\rangle $ is antisymmetric with respect to the sign reversal of $\lambda $,
while beyond the second-order boundary (dotted) $\langle \hat{p}^{2}\rangle $/$\langle \hat{x}^{2}\rangle $ is
vanishing relatively in $\lambda >0$ regime. The
ratio of $\langle \hat{x}^{2}\rangle $ and $\langle \hat{p}^{2}\rangle $ is
reversed in $\lambda <0$ regime, with a first-order boundary (long-dashed) at $\lambda =0$. This scenario forms a tricritical point
which is moving toward smaller-coupling direction for a positive $\chi $ and
toward larger-coupling\ direction for a negative $\chi $ as compared with
the case at $\chi =0$ marked by the dot in Figure \ref{fig-diagrams-low-w}a,c.

\subsubsection{Adiabatic Boundary}

In the absence of the nonlinear Stark coupling,
$\left\vert x_{m}^{A}\right\vert $ never goes beyond bottom of the bare
potential $v_{\sigma _{z}}$, i.e., the displacement renormalization ratio $\zeta =
x_{m}^{A}/x_{z,y}$ as indicated in Figure \ref{fig-Vari-eSC}e, where $x_{z}=g_{z}^{\prime }$ for $\lambda >0$ and $x_{z}=g_{y}^{\prime }$ for $\lambda <0$, is always smaller than $1$.\cite{Ying2015,Ying-2021-AQT} Now in the presence of the Stark
coupling, it may be more favorable to go farther away from the origin, and
finally beyond the potential bottom at a boundary
\begin{eqnarray}
\left\vert \lambda _{c}^{\zeta 1}\right\vert  &=&-1+2\sqrt{\frac{2\left(
1-\chi ^{2}\right) }{\chi \left( 3+\chi ^{2}\right) }}\frac{g_{\mathrm{s}}}{g%
}, \\
\left\vert \lambda _{c}^{\zeta 2}\right\vert  &=&-1+2\sqrt{\frac{2}{\left(
-\chi \right) }}\frac{g_{\mathrm{s}}}{g}, \\
g_{c}^{\zeta 1} &=&\frac{2g_{\mathrm{s}}}{1+\left\vert \lambda
\right\vert }\sqrt{\frac{2\left( 1-\chi ^{2}\right) }{\chi \left( 3+\chi
^{2}\right) }},  \label{gC-adiabtic-1} \\
g_{c}^{\zeta 2} &=&\frac{2g_{\mathrm{s}}}{1+\left\vert \lambda
\right\vert \ }\sqrt{\frac{2}{\left( -\chi \right) }},
\label{gC-adiabtic-2}
\end{eqnarray}%
for $\chi >0$ and $\chi <0,$ respectively. We show this boundary by the dot-dashed lines in Figure \ref{fig-diagrams-low-w}a,c,e, as compared with the numeric boundary of $\text{sign}[1-|\langle \hat{x}^{2}\rangle -\langle \hat{p}^{2}\rangle| /x_{z,y}^2]$ (note here $\langle \hat{p}^{2}\rangle$ is negligible in $\lambda >0$ regime as later on proven in Section \ref{Sect-sym-breaking}). Along this boundary the effective
particle keeps staying at the potential bottom position, which is adiabatic in the sense that the particle
is always following the potential.

\subsubsection{Vanishing-$\langle
\sigma _{x}\rangle $  Boundary and Coincidence with Adiabatic boundary}

As mentioned in Section \ref{Sect-Model}, $\langle
\sigma _{x}\rangle $ reflects flipping in the spin space and tunneling in the spatial space. Figure \ref{fig-diagrams-low-w}b,d show $\langle
\sigma _{x}\rangle $ which, unlike $\langle \hat{x}^{2}\rangle -\langle \hat{%
p}^{2}\rangle $, is symmetric with respect to $\lambda $. From the phase
diagrams of $\langle \sigma _{x}\rangle $ we find another phase boundary in $\lambda <0$ regime,
as plotted by the dashed lines in Figure \ref{fig-diagrams-low-w}d,f, which separates the positive and negative regimes
of $\langle \sigma _{x}\rangle $ at
\begin{eqnarray}
\left\vert \lambda _{c}^{\sigma _{x}}\right\vert  &=&-1+2\sqrt{\frac{2}{%
\left( -\chi \right) }}\frac{g_{\mathrm{s}}}{g},  \label{LambdaC-SpinX} \\
g_{c}^{\sigma _{x}} &=&\sqrt{\frac{2}{\left( -\chi \right) }}\frac{2g_{%
\mathrm{s}}}{\left\vert \lambda \right\vert +1\ },  \label{gc-SpinX} \\
\chi _{c}^{\sigma _{x}} &=&-\frac{8}{\left( \left\vert \lambda \right\vert
+1\right) ^{2}\left( g^{2}/g_{\mathrm{s}}^{2}\right) },  \label{ChiC-SpinX}
\end{eqnarray}%
as extracted by $\langle \sigma _{x}\rangle =0$ (See analytic expression of $%
\langle \sigma _{x}\rangle $ in Eq. (\ref{SpinX-Lambda-Scaling})).

Comparing Figure \ref{fig-diagrams-low-w}e,f one may notice the vanishing-$\langle \sigma _{x}\rangle $ boundary coincides with the
second adiabatic boundary, $g_{c}^{\sigma _{x}}=g_{c}^{\zeta 2}$, as also confirmed by Equations (\ref{gc-SpinX}) and (\ref{gC-adiabtic-2}). In fact,
under a negative $\chi $ the Stark-coupling energy is counteracting with the
tunneling energy, as one can see from the $\Omega $ term and the $x^{2}$\
term in Equation (\ref{H-SemiClassical}). The vanishing-$\langle \sigma _{x}\rangle $
boundary marked by different parameters in (\ref{LambdaC-SpinX})-(\ref{ChiC-SpinX}) is the point where
the Stark coupling energy $E_{\Omega }=\frac{\Omega }{2}\langle \sigma _{x}\rangle $ and the tunneling energy
$E_{\chi }=\chi \omega \langle \hat{n}\sigma _{x}\rangle $ are canceling. Indeed at
this point, not only the expectation of spin flipping $\sigma _{x}$ itself
is vanishing, but also the effective coefficients of $\sigma _{x}$ cancel:
\begin{equation}
\langle \sigma _{x}\rangle =0, \ \ \ \frac{\Omega }{2}+\frac{\chi \omega }{2}x^{2}=0,
\end{equation}%
at $x=x_{m}^{A}$ and $g=g_{c}^{\sigma _{x}}$. Consequently the Stark
coupling and tunneling term does not come to effect here and only the bare
potential $v_{\sigma _{z}}$ play the role.

Besides realizing that the sign reversal of $\langle \sigma _{x}\rangle $ only occurs in the
negative-$\chi $ regime, as indicated by the square root $\sqrt{-2/\chi }$ in
(\ref{LambdaC-SpinX}), we also see that the two boundaries $g _{c}^{(\lambda ,\chi )}$
and $g _{c}^{\sigma _{x}}$ does not meet unless at $\chi =-1$, as demonstrated by the Figure \ref{fig-diagrams-low-w}f.

\begin{figure*}[tbph]
\centering
\includegraphics[width=2.0%
\columnwidth]{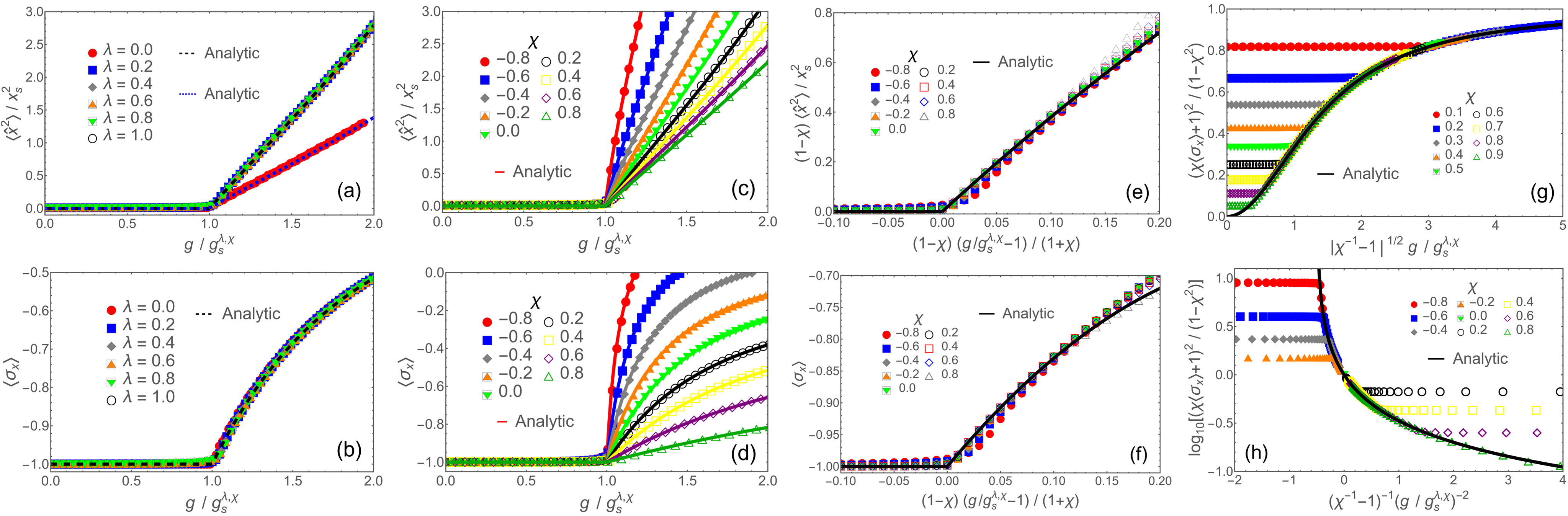}
\caption{\textit{Different levels of scaling relations in the low-frequency
limit.} (a,b) various $\lambda$ at a fixed $\chi$. (c-h)
various $\chi$ at a fixed $\lambda$. (a) $\langle \hat{x}^2
\rangle/x_{\mathrm{s}}^2$ versus $g/g_{\mathrm{s}}^{\lambda,%
\chi}$ at $\chi=0.4$. (b) $\langle \sigma_x \rangle$ versus $%
g/g_{\mathrm{s}}^{\lambda,\chi}$ at $\chi=0.4$. (c) $%
\langle \hat{x}^2 \rangle/x_{\mathrm{s}}^2$ versus $g/g_{\mathrm{s}}^{%
\lambda,\chi}$ at $\lambda=0.5$. (d) $\langle
\sigma_x \rangle$ versus $g/g_{\mathrm{s}}^{\lambda,%
\chi}$ at $\lambda=0.5$. (e) $(1-\chi)\langle \hat{x}^2
\rangle/x_{\mathrm{s}}^2$ versus $(1-\chi)(g/g_{\mathrm{s}}^{%
\lambda,\chi}-1)/(1+\chi)$ around transition at $%
\lambda=0.5$. (f) $\langle \sigma_x \rangle$ versus $(1-\chi%
)(g/g_{\mathrm{s}}^{\lambda,\chi}-1)/(1+\chi)$
around transition at $\lambda=0.5$. (g) $(\chi\sigma%
_x+1)^2/(1-\chi ^2)$ versus $|\chi^{-1}-1|^{1/2}g/g_{\mathrm{%
s}}^{\lambda,\chi}$ at $\lambda=0.5$. (h) $\log
_{10}[(\chi\sigma_x+1)^2/(1-\chi ^2)]$ versus $|%
\chi^{-1}-1|^{-1}(g/g_{\mathrm{s}}^{\lambda,\chi%
})^{-2}$ at $\lambda=0.5$. Here $\omega =0.01\Omega$ for
(a,b,c,d,g,h) and $\omega =0.001\Omega$ for (e,f)}
\label{fig-scaling}
\end{figure*}

\subsection{Scaling Relations at Fixed Stark Couplings}

\label{Sect-Scaling-fixed-Stark}

There exist some scaling relations for the critical behavior. The scaling
relation forms a universality of critical properties in the linear
anisotropic QRM,\cite{LiuM2017PRL,Ying-2021-AQT} while some more general
universality can be found in the presence of the nonlinear Stark coupling.
We first consider the case at fixed nonlinear Stark
couplings. In the low-frequency limit, after the transition at $g _{c}^{(\lambda ,\chi )}$ we see the
effective spatial position in $n$-order%
\begin{equation}
\frac{\langle \hat{x}^{n}\rangle }{2^{n/2}x_{\mathrm{s}}^{n}}=\left( -\frac{%
\overline{g}_{\lambda }^{2}+\chi }{\chi ^{2}}+\frac{\overline{g}_{\lambda }}{%
\chi ^{2}}\sqrt{\frac{\overline{g}_{\lambda }^{2}+2\chi }{1-\chi ^{2}}}%
\right) ^{n/2},  \label{xn-Lambda-Scaling}
\end{equation}%
for the positive-$\lambda $ regime and similarly $\langle \hat{p}^{n}\rangle
$ for the negative-$\lambda $ regime, while it is vanishing before the
transition. Here we have defined
$x_{\mathrm{s}}=\sqrt{2}g_{\mathrm{s}}/\omega $ and
$\overline{g}_{\lambda }=g/g_{c}^{\lambda }=\sqrt{1-\chi }\overline{g}_{\lambda ,\chi }$,
where
$\overline{g}_{\lambda ,\chi }=g/g_c^{\lambda ,\chi }$.
So, the scaled expression (\ref{xn-Lambda-Scaling})
is a function of $\overline{g}_{\lambda }$ or $\overline{g}_{\lambda ,\chi },
$ thus being universal for all values of $\lambda $. \textbf{Figure} \ref{fig-scaling}a illustrates
the numerical expectation $\langle \hat{x}^{2}\rangle $ for
different anisotropies at a given Stark coupling $\chi =0.4$ for a low
frequency $\omega =0.01\Omega $, one sees that with the scaling $%
g/g_{c}^{\lambda ,\chi }$ for the coupling strength and $\langle \hat{x}%
^{2}\rangle /x_{\mathrm{s}}^{2}$ for the quadratic position all data of
different anisotropies collapse into a single line, except the JC case $%
\lambda =0$. The analytic scaling relation (\ref{xn-Lambda-Scaling}) is
plotted by the black dashed line which coincides with the numerics.

We also find the critical scaling relation for the spin expectation
\begin{equation}
\langle \sigma _{x}\rangle =\frac{1}{\chi }\left( -1+\overline{g}_{\lambda }%
\sqrt{\frac{1-\chi ^{2}}{2\chi +\overline{g}_{\lambda }^{2}}}\right)
\label{SpinX-Lambda-Scaling}
\end{equation}%
which is also independent of $\lambda$ and agrees well with the numerical scaling data,
as shown in Figure \ref{fig-scaling}b where all anisotropy cases collapse into a single line,
including $\lambda =0$. We see that the scaling of $\langle \sigma
_{x}\rangle $ with respect to the anisotropy is more universal than $\langle
\hat{x}^{n}\rangle ,\langle \hat{p}^{n}\rangle $ in the sense there is no
discontinuity at the point $\lambda =0$ in $\langle \sigma _{x}\rangle $.
The difference comes from the fact that $\langle \hat{x}^{n}\rangle $ and $%
\langle \hat{p}^{n}\rangle $ suffer from a spontaneous symmetry breaking of
the duality exchange \cite{Ying-2021-AQT} while $\sigma _{x}$ remains
unaffected in the duality exchange.

\subsection{Symmetry Breaking and Singularity in the $\langle \hat{x}%
^{2}\rangle $-Scaling Relation around $\lambda =0$}

\label{Sect-sym-breaking}

As seen from $H_x$ and $H_p$ in Section \ref{Sect-Model} the $\lambda =0$ case has an $x$-$p$
duality symmetry which is however broken once a non-zero value of $\lambda $
is introduced. It is this symmetry breaking that leads to the singular
behavior of $\lambda =0$ in the scaling relation of $\langle \hat{x}%
^{2}\rangle $ afore-mentioned in Sec. \ref{Sect-Scaling-fixed-Stark}. Here
we shall clarify the mechanism more explicitly.

For the $\lambda =0$ case, from the exact wavefunction (\ref{WaveF-JC-n}) we
see the expectations are indeed equal in accordance with the $x$-$p$ duality
symmetry, $\langle \hat{x}^{2}\rangle =\langle \hat{p}^{2}\rangle =[(n+\frac{%
1}{2})C_{n\Uparrow }^{\left( \pm \right) 2}+(n+\frac{3}{2})C_{n\Downarrow
}^{\left( \pm \right) 2}]/N$ with $n$ substituted by $n_{\min }$ in (\ref%
{nMin}), which approach to
\begin{equation}
\frac{\langle \hat{x}^{2}\rangle _{\lambda =0}}{x_{\mathrm{s}}^{2}}=\frac{%
\langle \hat{p}^{2}\rangle _{\lambda =0}}{x_{\mathrm{s}}^{2}}=-\frac{%
\overline{g}_{\lambda }^{2}+\chi }{\chi ^{2}}+\frac{\overline{g}_{\lambda }}{%
\chi ^{2}}\sqrt{\frac{\overline{g}_{\lambda }^{2}+2\chi }{1-\chi ^{2}}}
\label{x2-JC}
\end{equation}%
in the low-frequency limit. Figure \ref{fig-scaling}a shows the agreements
of analytic result (blue dotted line) of (\ref{x2-JC}) and the numerics (red
dots).

Once away from the $\lambda =0$ line, the symmetry breaking leads to an
imbalance of $\langle \hat{x}^{2}\rangle $ and $\langle \hat{p}^{2}\rangle $%
, as we can see from a comparison of the energy $H_{\mathrm{SC}}^{x}$ and $%
H_{\mathrm{SC}}^{p}$ in (\ref{H-SemiClassical}). In fact, the minimized
energy of $H_{\mathrm{SC}}^{x}$ and $H_{\mathrm{SC}}^{p}$ as in (\ref{Ea-SC}%
) can be unified to be a same function of $g_{zy}^{\prime }$
\begin{eqnarray}
E_{\mathrm{SC}}^{A}(g_{zy}^{\prime }) &=&-\frac{1}{2}g_{zy}^{\prime 2}\omega
-\frac{g_{zy}^{\prime 2}\left( 2-\chi ^{2}\right) \omega +\chi \Omega }{%
2\chi ^{2}}  \nonumber \\
&&+\frac{g_{zy}^{\prime }\left( 1-\chi ^{2}\right) \omega }{\chi ^{2}}\sqrt{%
\frac{g_{zy}^{\prime 2}+\chi \frac{\Omega }{\omega }}{1-\chi ^{2}}},
\end{eqnarray}%
with $g_{zy}^{\prime }=g_{z}^{\prime }$ for $H_{\mathrm{SC}}^{x}$ and $%
g_{zy}^{\prime }=g_{y}^{\prime }$ for $H_{\mathrm{SC}}^{p}$. $E_{\mathrm{SC}%
}^{A}$ as a function of $g_{zy}^{\prime }$ has a maximum point $g_{zy,\max
}^{\prime }$ at
\begin{equation}
g_{zy,\max }^{\prime }=g_{zy,c}^{\prime }=\sqrt{\left( 1-\chi \right) \Omega
/\left( 2\omega \right) }
\end{equation}%
which happens to be the critical point $g_{zy,c}^{\prime }$ for the
transition as decided by $E_{\mathrm{SC}}^{A}=E_{\mathrm{SC}}^{B}$. After
the transition $E_{\mathrm{SC}}^{A}$ becomes a decreasing function of $%
g_{zy}^{\prime }$, which can be can be seen from the derivative $\partial E_{%
\mathrm{SC}}^{A}/\partial g_{zy}^{\prime }=d_{A}-d_{B}$, where $%
d_{A}^{2}-d_{B}^{2}=-C_{d}\left( g_{zy}^{\prime 2}-g_{zy,\max }^{\prime
2}\right) [g_{zy}^{\prime 2}+\left( 1+\chi \right) \Omega /\left( 2\omega
\right) ]$ and $C_{d}=4\omega ^{2}/[\chi ^{2}(g_{zy}^{\prime 2}+\chi \frac{%
\Omega }{\omega })]$, being negative after $g_{zy,\max }^{\prime }$. Thus, a
larger value of $g_{zy}^{\prime }$ provided by $g_{z}^{\prime }$ and $g_{y}^{\prime }$ will be more favorable for the candidate of the GS.

Note in
the positive-$\lambda $ regime $g_{z}=\frac{\left( 1+\lambda \right) }{2}g$
has a larger value than $g_{y}=\frac{\left( 1-\lambda \right) }{2}g$,
consequently $H_{\mathrm{SC}}^{x}$ provides a lower energy than $H_{\mathrm{%
SC}}^{p}$ due to the decreasing function $E_{\mathrm{SC}}^{A}(g_{zy}^{\prime
})$. As a result, the GS from $H_{\mathrm{SC}}^{x}$ has a vanishing $%
\langle \hat{p}^{2}\rangle $ but a finite $\langle \hat{x}^{2}\rangle $ in (%
\ref{xn-Lambda-Scaling}) that is twice of $\langle \hat{x}^{2}\rangle
_{\lambda =0}$ in (\ref{x2-JC}), as in the contrast displayed in Figure \ref%
{fig-scaling}a. Reversely in the negative-$\lambda $ regime, $\langle \hat{x}%
^{2}\rangle $ is vanishing and $\langle \hat{p}^{2}\rangle $ is finite as $%
g_{y}$ is larger than $g_{z}$.

From the above discussion we see that the singular behavior in the $\langle
\hat{x}^{2}\rangle $-scaling relation is a consequence of the difference from half
weights of $\langle \hat{x}^{2}\rangle $ and $\langle \hat{p}^{2}\rangle $ for $%
\lambda =0,$\ full weight of $\langle \hat{x}^{2}\rangle $ for $\lambda >0$,
and full weight of $\langle \hat{p}^{2}\rangle $ for $\lambda <0$, in the $x$%
-$p$ duality symmetry breaking. We can get rid of the singular behavior and
get a more unified scaling relation%
\begin{equation}
\frac{\langle \hat{x}^{2}\rangle +\langle \hat{p}^{2}\rangle }{x_{\mathrm{s}%
}^{2}}=2\left( -\frac{\overline{g}_{\lambda }^{2}+\chi }{\chi ^{2}}+\frac{%
\overline{g}_{\lambda }}{\chi ^{2}}\sqrt{\frac{\overline{g}_{\lambda
}^{2}+2\chi }{1-\chi ^{2}}}\right) ,
\end{equation}
which holds for any anisotropy including $\lambda =0$, as shown in Figure \ref{fig-diversity}a.

\subsection{Local Scaling Relations for Various Stark Couplings around
Transition}

\label{Sect-Scaling-all-Stark}

In the last two sections we have extended the scaling relation with respect to anisotropy from the absence to the presence of nonlinear Stark coupling.
A more general scaling relation would be universal not only for all
anisotropies but also for various Stark couplings. The $\lambda $-universal
scaling relations (\ref{xn-Lambda-Scaling}) and (\ref{SpinX-Lambda-Scaling})
are however not unified for different values of $\chi $, as shown in Figure \ref{fig-scaling}c,d. A general scaling relation
universal for both $\lambda $ and $\chi $ holding globally for any strength of
coupling is not readily available. Nevertheless, since critical exponent
depends on the behavior in the vicinity of the transition, we can extract
some scaling relations around the transition by expansion%
\begin{eqnarray}
&&\frac{\left( 1-\chi \right) \langle \hat{x}^{2}\rangle }{2\hat{x}_{\mathrm{%
s}}^{2}}=2d\overline{g}_{\lambda ,\chi ,\omega }-d\overline{g}_{\lambda
,\chi ,\omega }^{2}+O(d\overline{g}_{\lambda ,\chi ,\omega }^{3})
\label{x2-Lambda-Chi-Scaling} \\
&&\langle \sigma _{x}\rangle =-1+2d\overline{g}_{\lambda ,\chi ,\omega }-3d%
\overline{g}_{\lambda ,\chi ,\omega }^{2}+O(d\overline{g}_{\lambda ,\chi
,\omega }^{3})  \label{SpinX-Lambda-Chi-Scaling}
\end{eqnarray}%
where $d\overline{g}_{\lambda ,\chi ,\omega }=\frac{1-\chi }{1+\chi }\left(
\frac{g}{g_{c}^{\lambda ,\chi }}-1\right) $, which are pure functions of $d%
\overline{g}_{\lambda ,\chi ,\omega }$ independent of $\lambda ,\chi ,\omega
$ in the first two orders. We present a comparison of the analytic scaling (%
\ref{x2-Lambda-Chi-Scaling}) and (\ref{SpinX-Lambda-Chi-Scaling}) with the
numerical data in Figure \ref{fig-scaling}e,f. The comparison shows that the
scaling relations basically hold for different values of $\chi $ indeed,
except near the unphysical limit $\chi = 1$ due to the
singular third-order term which takes the form of $2d\overline{g}_{\lambda
,\chi ,\omega }^{3}(2-3\chi )/(1-\chi ).$

\subsection{Global Scaling Relation for Various Stark Couplings after
Transition}

Still, a more robust scaling relation can be obtained for $\langle \sigma _{x}\rangle$, universal for $\lambda ,\chi $ and low frequencies without
limitation of $\chi $ or the critical regime around the transition. In fact
we find that the following scaling relation after
the transition
\begin{equation}
\frac{\left( \chi \langle \sigma _{x}\rangle +1\right) ^{2}}{1-\chi ^{2}}=%
\frac{1}{2\chi \overline{g}_{\lambda }^{-2}+1}=\frac{1}{2[\left( \chi
^{-1}-1\right) \overline{g}_{\lambda ,\chi }^{2}]^{-1}+1}.
\label{SpinX-Scaling-All}
\end{equation}%
Figure \ref{fig-scaling}g shows the scaling relation (\ref{SpinX-Scaling-All}) as a function of $\sqrt{\left(
\chi ^{-1}-1\right) }\overline{g}_{\lambda ,\chi }$. Here in the figure the
horizontal symbols are the numeric data before the phase transition, while
in the critical regime after the phase transition all data in different
values of $\chi $ collapse into a same line which coincides with the
analytic scaling (\ref{SpinX-Scaling-All}). Note here the values of $\chi $
are positive, while negative $\chi $ also has a similar scaling behavior but
in a different branch. Nevertheless both negative and positive $\chi $ can
be finally unified in a scaling as a function of $\left( \chi ^{-1}-1\right)
^{-1}\overline{g}_{\lambda ,\chi }^{-2}$, as in Figure \ref{fig-scaling}h.
Note that the variation in $\lambda $ has been scaled without any limitation
as shown both numerically and analytically in Sec. \ref%
{Sect-Scaling-fixed-Stark}, thus the scaling (\ref{SpinX-Scaling-All}) is
valid for both $\lambda $ and $\chi $.

These scaling relations indicate that the properties in the critical regime
obey a universal law, despite that the parameters are different in the
anisotropy, the Stark coupling ratio and the frequencies (in low
frequencies).

\begin{figure*}[t]
\centering
\includegraphics[width=1.65%
\columnwidth]{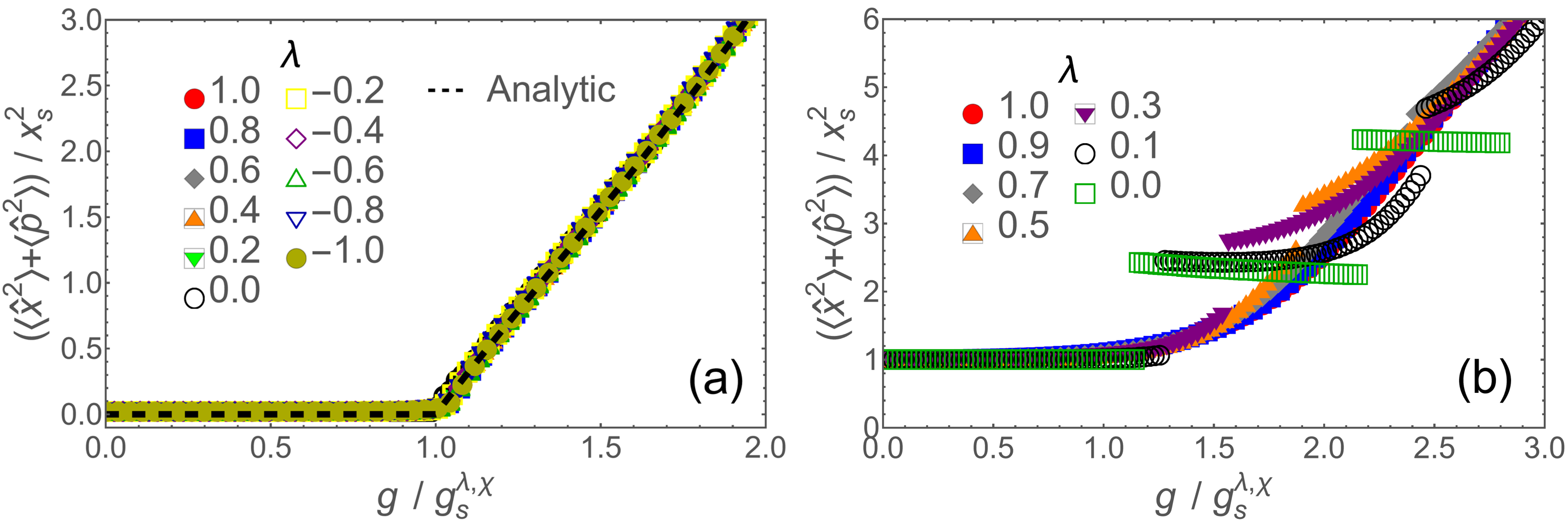}
\caption{\textit{Breaking down of the critical universality at finite
frequencies.} (a) Critical scaling relation for $\langle\hat{x}^2\rangle
+\langle\hat{p}^2\rangle$ universal for all anisotropy ratios $%
\lambda$ at a low frequency $\omega = 0.01 \Omega$. (b) Breaking
down of the critical universality at a finite frequency $\omega =
0.5 \Omega$. Here $\chi =0.2 $ in (a,b). }
\label{fig-diversity}
\end{figure*}

\section{Breaking Down of Critical Universality at Finite Frequencies}

\label{Sect-Diversity}

As addressed in Secs. \ref{Sect-Scaling-fixed-Stark}-\ref{Sect-Scaling-all-Stark}, all levels of the scaling relations are valid
under the condition of low-frequency limit. At finite frequencies the critical
universality will break down and different scenarios arise. Indeed, as
illustrated in \textbf{Figure} \ref{fig-diversity}b at a finite frequency $\omega
=0.5\Omega $ for a fixed Stark coupling $\chi =0.2$, the expectations $\langle \hat{x}^{2}\rangle +\langle \hat{p}^{2}\rangle $
in different ratios of anisotropy are not
collapsing into a single line any longer. We see that $\langle \hat{x}^{2}\rangle +\langle \hat{p}^{2}\rangle $ in different $\lambda $ not only increases in
various gradients but also fragment into disconnected sections. At the
section breakings actually emerging are a series of first-order phase
transitions. Note here that fixing the Stark coupling as in Figure \ref%
{fig-scaling}a,b is the lowest level of scaling. Now even the lowest level
of scaling relation has broken down, not to mention the collapse of the
higher levels of scaling relations under both various anisotropies and
different nonlinear Stark couplings in Figure \ref{fig-scaling}e-h. Thus we
see the critical universality collapses and the systems properties are
dominant by diversity which, opposite to universality with common feature,
is the quality to be diverse or different.

\begin{figure*}[tbph]
\centering
\includegraphics[width=2.0%
\columnwidth]{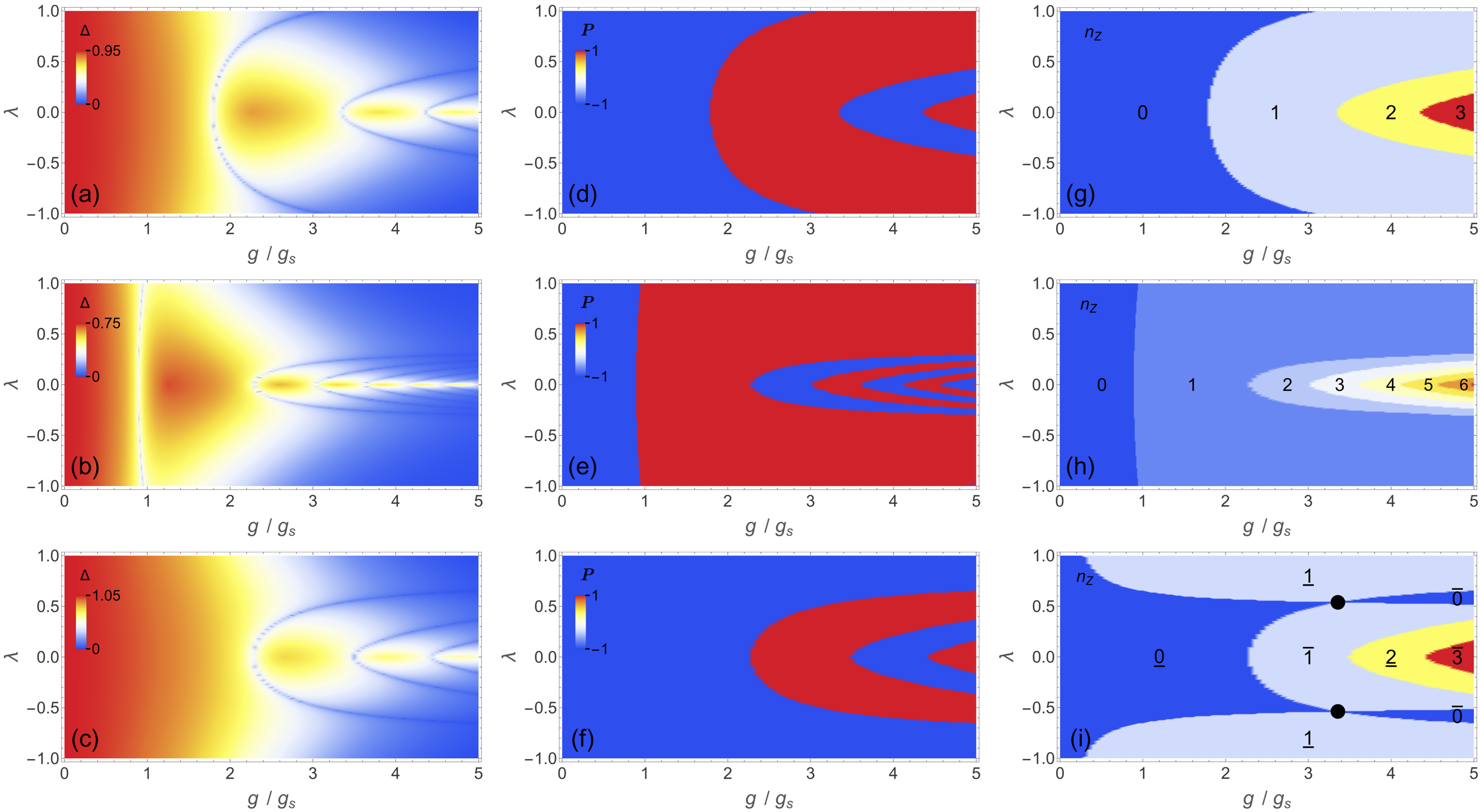}
\caption{\textit{Topological classification and topological quadruple point
at a finite frequency $\omega =0.5 \Omega$ with fixed Stark
couplings.} The first excitation gap $\Delta$ (a-c), the GS parity
$P$ (d-f), and the GS node number $n_Z$ (g-i) at fixed Stark
couplings $\chi =0.2$ (a,d,g), $\chi =0.8$ (b,e,h), and $%
\chi =-0.3$ (c,f,i). The numbers in (g-i) are $n_Z$, while the
overlines and underlines in (i) represent negative and positive parities.
The black dots mark the topological quadruple points \{$g_{\mathrm{TQ}}$, $%
\lambda_{\mathrm{TQ}}$ \} (\ref{g-TQ-fix-Stark}). }
\label{fig-topo-fix-gStark}
\end{figure*}

\begin{figure*}[tbph]
\centering
\includegraphics[width=2.0%
\columnwidth]{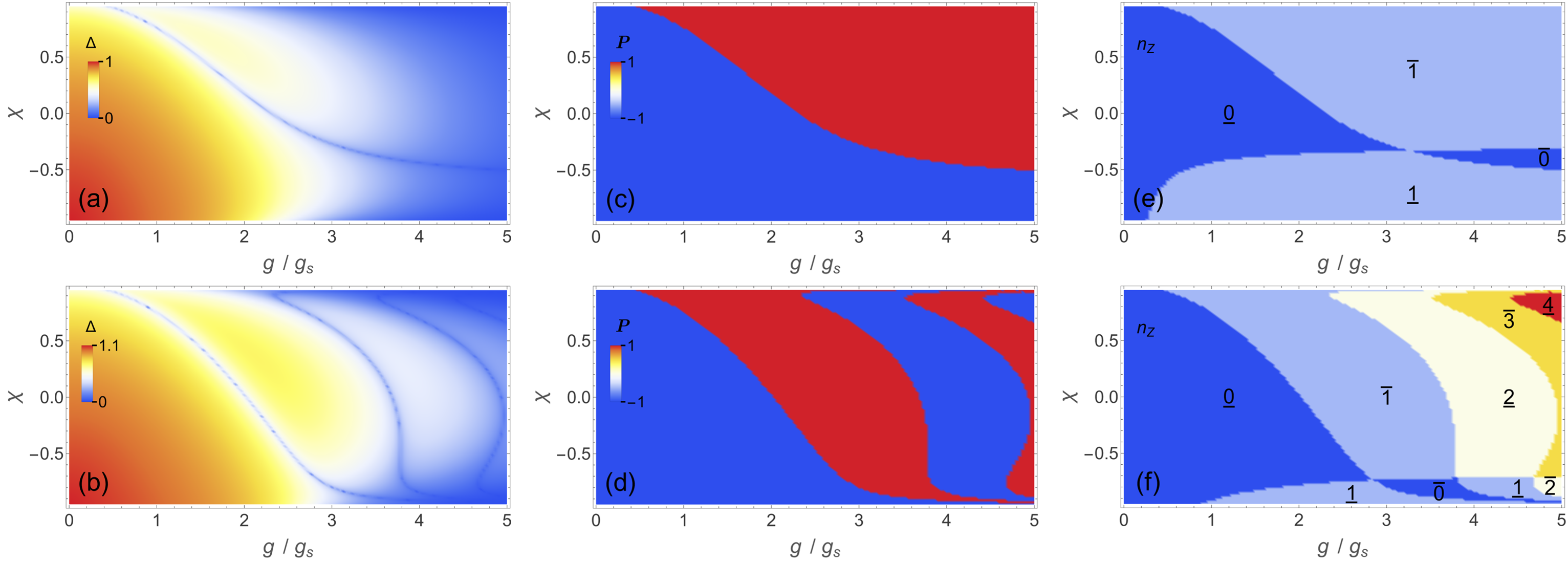}
\caption{ \textit{Topological classification and topological quadruple
points at a finite frequency $\omega =0.5 \Omega$ with fixed
anisotropy strengths.} The first excitation gap $\Delta$ (a,b), the
GS parity (c,d), and the GS node number (e,f) at $%
\lambda =0.5$ (a,c,e), $\lambda =0.15$ (b,d,f). }
\label{fig-topo-fix-lambda}
\end{figure*}

\begin{figure*}[tbph]
\centering
\includegraphics[width=2.0%
\columnwidth]{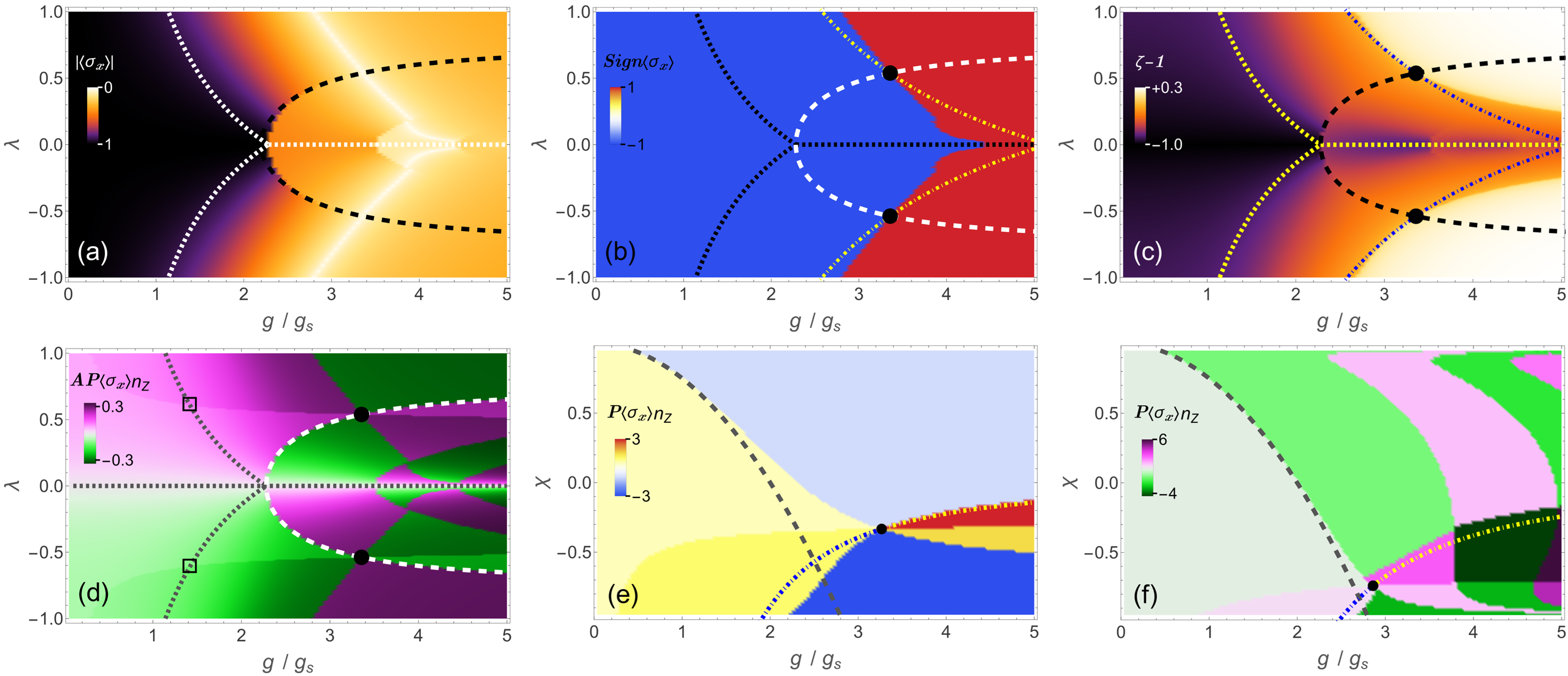}
\caption{\textit{Composite phase diagrams: multicriticality, topological
quadruple points, composite hexaple points, and $\langle \sigma_x
\rangle$-invariant points.} a) Amplitude of $\langle \sigma %
_{x}\rangle $ in $g$-$\lambda$ plane, plotted by $| \langle %
\sigma _{x}\rangle | ^{1/2}$. b) Sign of $\langle \sigma _{x}\rangle$%
. c) $\zeta -1$, with amplitude amplified by $| \zeta -1 |
^{1/3}$. d) Composite phase diagram of $B_1=(1+n_Z)^{1/4} \text{sign}(\langle
\sigma _{x}\rangle) P A$. Here $\chi=-0.3$ in a-d), $P$ is
the parity and $A=\langle a^{\dagger}a^{\dagger}\rangle/A_0$ as defined in
Figure \ref{fig-diagrams-low-w}. e) $B_2=(1+n_Z)P[1/2+\text{sign}(\langle
\sigma _{x}\rangle)]$ in $g$-$\chi$ plane at $\lambda%
=0.5$. f) $B_2$ at $\lambda=0.15$. $\omega=0.5\Omega$ in all
panels. The dotted lines in (a-d) are $g_c^{\lambda,\chi}$,
while horizontal ones are symmetric lines at $\lambda=0$. The dashed
lines are $g_{\mathrm{T1}}^{\lambda,\chi}$ (\ref%
{gT1-Polaron}) in (a) and $g_{\mathrm{T1,E}}^{\lambda,\chi}$
(\ref{gT1-Exact}) in (b-d). In (b-f) the dot-dashed lines are $%
g_c^{\sigma_x}=g_c^{\zeta 2}$ (\ref{gc-SpinX},%
\ref{gC-adiabtic-2}) in low-frequency limit and the black dots mark
\{$g_{\mathrm{TQ}}$, $\lambda_{\mathrm{TQ}}$ or $\chi_{%
\mathrm{TQ}}$ \} (\ref{g-TQ-fix-Stark},\ref{g-TQ-fix-Lambda}%
). The dashed lines in (e-f) are $g_c^{\mathrm{JC-Stark}}$ (\ref%
{gc-JC-Stark}). }
\label{fig-diagrams-composite}
\end{figure*}

\section{Topological Classification at Finite Frequencies}
\label{Sect-Top-Classification}

Although the critical universality in low frequencies breaks down at finite
frequencies and the properties are diversified, among the diversity we can
still extract some common feature but from the topological structure of the
GS wave function.\cite{Ying-2021-AQT,Ying-gapped-top} Indeed,
within each emerging phase at finite frequencies the GS wave
function has a same node number, i.e. the number of zeros $n_{Z}$. Wave
functions with different node numbers are topologically different in the
sense that, by fixing a node number $n_{Z}$, one cannot go to another $n_{Z}$
state by continuous shape deformation of the wave function, just as one
cannot change a torus into a sphere by a continuous deformation which is a
well-known illustration for topological difference.
Nodes of polynomial functions are also related to topological Galois theory in connecting algebra to topology.\cite{topo-Galois-theory}
The node number is the same universally for all system parameters within a phase. This leads us to a
topological classification which is not only valid for the linear
anisotropic QRM \cite{Ying-2021-AQT,Ying-gapped-top} but also in the
presence of the nonlinear Stark coupling as shown in the following. On the other hand, nonlinear coupling
will lead us to new phenomena unexpected in linear coupling, such as topological quadruple points, composite sextuple points and $\langle \sigma _{x}\rangle$-invariant points.

\subsection{Conventional Topological Transitions with Gap Closing}

\label{Sect-Top-with-gap-closing}

\textbf{Figure} \ref{fig-topo-fix-gStark}a-c show the first excitation gap $\Delta $
in the $g$-$\lambda $ plane for $\omega =0.5\Omega $ at a fixed Stark
coupling ratio $\chi =0.1$,\ $0.4$, $-0.3$. We see that some series of
boundaries emerge where the gap is actually closing and re-opening. Figure %
\ref{fig-topo-fix-gStark}d-f show the phase diagrams of parity
correspondingly, with the negative and positive parities represented by the
colors in blue and red, respectively. Comparing Figure \ref%
{fig-topo-fix-gStark}d-f with Figure \ref{fig-topo-fix-gStark}a-c we see
that the parity is reversed at the gap closing boundaries. Note the parity
has only two values which are not enough to distinguish the series phases
that emerge with the series of transitions. Something beyond the parity is
needed to understand the nature of the transitions, which turns out to be
topological structure of the wave function. Indeed, the node number $n_{Z}$
of the GS wave function changes across each boundary of the gap
closing and parity reversal, as shown by Figure \ref{fig-topo-fix-gStark}g-i
where the numbers mark $n_{Z}$ of different phases. These transitions are
analogs of the conventional TPTs that occurs at gap closing without symmetry
breaking.\cite{Ying-2021-AQT}

\subsection{Unconventional Topological Transitions without Gap Closing}

\label{Sect-Top-without-gap-closing}

Besides the conventional TPTs with gap closing, unconventional TPTs may also
occur without gap closing. Figure \ref{fig-topo-fix-gStark}i shows the phase
diagram of node number at a negative Stark coupling ratio $\chi =-0.3$. We
see that, besides the transitions at the gap closing and parity reversal
corresponding to Fig. \ref{fig-topo-fix-gStark}c,f, there are two boundaries
that have no match of either gap closing or parity change. \textbf{Figure} \ref{fig-topo-fix-lambda}
shows the phase diagrams in the $g$-$\chi $ plane at fixed anisotropy
strengths $\lambda =0.15\ $(a,c,e) and $\lambda =0.5\ $(b,d,f). We see that
besides the conventional transitions with gap closing and parity reversal, a transition boundary of
node number without gap closing is also showing up in the negative-$\chi $
regime.

These additional transitions are analogs of the unconventional TPTs without gap closing
in condensed matter which may occur in some particular situations, such as
in the presence of a strong electron-electron interaction in the quantum
spin Hall effect \cite{Amaricci-2015-no-gap-closing} or in the presence of
disorder with Berry curvature separation in the quantum anomalous Hall
effect.\cite{Xie-QAH-2021}

\subsection{Topological Quadruple Points}

\label{Sect-Top-Quadruple}

\begin{figure*}[tbph]
\centering
\includegraphics[width=2.0\columnwidth]{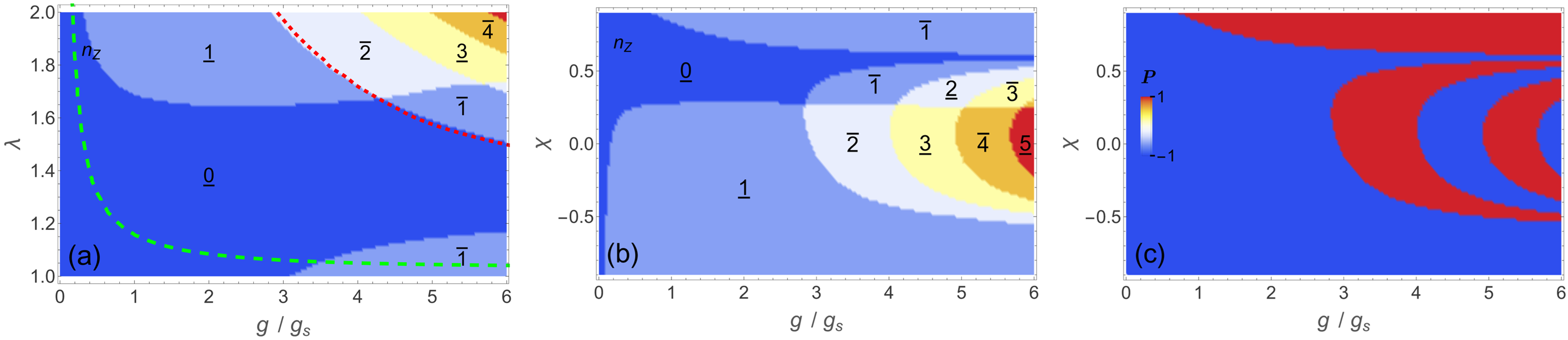}
\caption{\textit{Phase diagrams in large-$\lambda$ regime at $%
\omega = 0.5\Omega $.} a) $n_Z$ in $g$-$\lambda$ plane at $%
\chi=0.1$. b) $n_Z$ in $g$-$\chi$ plane at $\lambda %
=2.0$. c) $P$ at $\lambda =2.0$. The numbers mark $n_Z$ with
underlines and overlines representing negative and positive parities. The dashed line and dotted line in (a) are respectively the $\underline{0}$/$\underline{1}$ and  $\underline{1}$/$\overline{2}$ boundaries at $\chi =0$.   }
\label{fig-large-lambda}
\end{figure*}
\begin{figure*}[tbph]
\centering
\includegraphics[width=2.0%
\columnwidth]{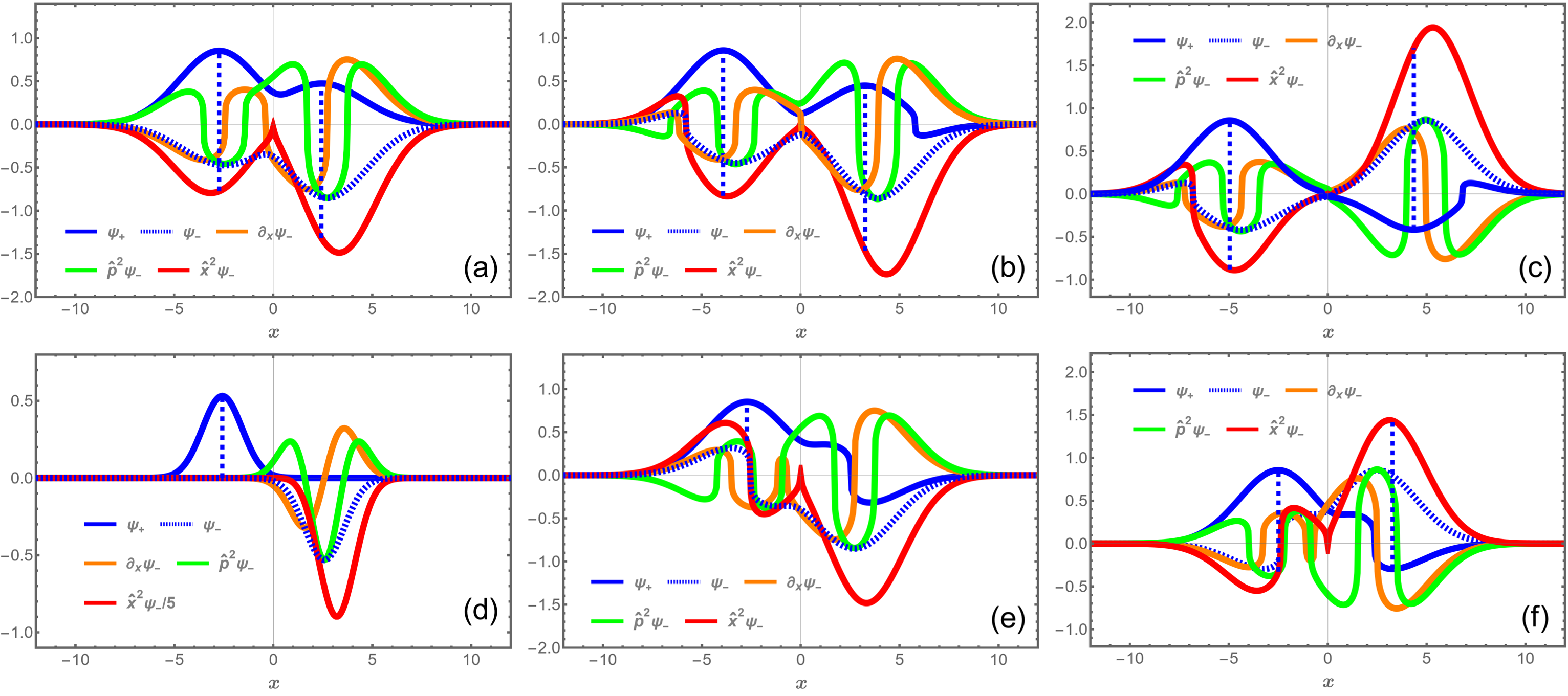}
\caption{ \textit{Mechanism analysis: GS wave function and energy
competitions.} $\psi_{+}$ (blue solid), $\psi_{-}$ (blue
dotted), $\partial_x\psi_{-}$ (orange), $\hat{p}^2\psi_{-}$
(green), $\hat{x}^2\psi_{-}$ (red) in $x$ space at a) $\chi%
=0.1$, $\lambda=1.1$, $g=2.6g_{\mathrm{s}}$ ($P=-1$, $n_Z=0$), b) $%
\chi=0.1$, $\lambda=2.0$, $g=2.6g_{\mathrm{s}}$ ($P=-1$ and $%
n_Z=1$), c) $\chi=0.1$, $\lambda=2.0$, $g=3.3g_{\mathrm{s}}$
($P=1$ and $n_Z=2$), d) $\chi=-0.3$, $\lambda=0.538$, $%
g=3.355g_{\mathrm{s}}$ ($P=-1$ and $n_Z=0$), e) $\chi=-0.3$, $%
\lambda=0.8$, $g=3.0g_{\mathrm{s}}$ ($P=-1$ and $n_Z=1$), f) $%
\chi=-0.3$, $\lambda=0.4$, $g=3.6g_{\mathrm{s}}$ ($P=1$ and $%
n_Z=1$). The vertical dashed lines mark the peak positions of $\psi%
_{+}$. In all panels $\omega=0.5\Omega$ and except (d) the amplitude
$A_m$ in each line is amplified by $A_m^{1/4}$ for better visibility.}
\label{fig-wavefunction}
\end{figure*}

In the absence of the nonlinear Stark coupling the unconventional TPTs occur
in $\lambda >1$ regime.\cite{Ying-gapped-top} Here we see that the nonlinear
Stark coupling is bringing the unconventional TPTs from the regime beyond
the QRM line ($\lambda =1$) to the intermediate regime $0<\lambda \leqslant 1
$ between the QRM and the JCM which are the most fundamental models in
light-matter interactions. The entering of the unconventional TPTs into the
intermediate regime has two consequences: On the one hand, the TPTs can
occur in the isotropic QRM with weak couplings, as indicated by Figure \ref{fig-topo-fix-gStark}i;
On the other hand, some topological quadruple points
are formed, as one finds in Figure \ref%
{fig-topo-fix-gStark}i and Figure \ref{fig-topo-fix-lambda}e,f.

Note that
the node number characterizes the topological structure of the wave function
within a same spin component, while the parity reflects the relative
structure between the two spin components. In the phase labels of Figure \ref{fig-topo-fix-gStark}i
and Figure \ref{fig-topo-fix-lambda}e,f we have
combined the node number and the parity to distinguish the phases better
from each other. We see that four topological phases meet at a topological
quadruple point, e.g. in Figure \ref{fig-topo-fix-gStark}i, marked by dots around which two phases have $%
n_{Z}=0$ and the other two have $n_{Z}=1$ while the parity is different for
the phases with a same node number. Such topological quadruple points mean
the boundary crossing of the conventional TPTs with gap closing
and unconventional TPTs without gap closing, which never happens for the
linear interaction in the absence of the nonlinear Stark coupling.\cite%
{Ying-gapped-top}

\subsection{Composite Phase Diagrams: Multicriticality, Composite Quadruple
Points, Composite Sextuple Points}

We have seen in Figure \ref{fig-topo-fix-gStark} that $P$ and $n_{Z}$, are
symmetric with respect to the sign reversal of $\lambda $ as it is also true
for $\langle \sigma _{x}\rangle ,\zeta $ discussed in Section \ref%
{Sect-phase-diagram-low-w}, while $\langle \hat{x}^{2}\rangle -\langle \hat{p%
}^{2}\rangle $ is antisymmetric in Figure \ref{fig-diagrams-low-w}a,c.
Here at a finite frequency $\zeta $ is extracted by ratio of the main-peak
position of $\psi _{\pm }\left( x\right) $ and the potential-bottom position
$g_{z}^{\prime }$ for $\lambda >0$ ($g_{y}^{\prime }$ for $\lambda <0$).
Combining these quantities for an overview will expose some underlying features.

\textit{Hexacritical point:} In \textbf{Figure} \ref{fig-diagrams-composite}d
we present a density plot for the composite quantity $(n_{Z}+1)^{1/4}P(\langle
\hat{x}^{2}\rangle -\langle \hat{p}^{2}\rangle )\text{sign}(\langle \sigma
_{x}\rangle )$ in the $g$-$\lambda $ plane at $\omega =0.5\Omega $ and $\chi
=-0.3$. Along the symmetric $\lambda =0$ line, one sees first a hexacritical
point around $g=2.3g_{\mathrm{s}}$ which is the crossing point of the
first-order boundary (white dashed line, see expressions in (\ref%
{gT1-Polaron},\ref{gT1-Exact})) and the second-order boundary $%
g_{c}^{\lambda ,\chi }$ (black dotted curves, (\ref{gC-SC})) as more
reflected by the amplitude of $\langle \sigma _{x}\rangle $ in Figure \ref%
{fig-diagrams-composite}a.

\textit{Composite quadruple/sextuple points:} Along the $\lambda =0$ line
following the afore-mentioned hexacritical point are a composite quadruple
point around $g=3.5g_{\mathrm{s}}$ and a composite sextuple point around $%
g=4.5g_{\mathrm{s}}$, while increasing $g$ one would see more composite
quadruple points beyond the plotting range. The composite sextuple point is
actually a quadruple point (not topological quadruple point) in $%
n_{Z}P(\langle \hat{x}^{2}\rangle -\langle \hat{p}^{2}\rangle )$ but the
sign-reversal boundary of $\langle \sigma _{x}\rangle $ renders it to be a
sextuple-like point. The $\langle \sigma _{x}\rangle $-sign-reversal
boundary as shown in Figure \ref{fig-diagrams-composite}b also leads to
another composite quadruple point away from the $\lambda =0$ line around $%
\{g/g_{\mathrm{s}},\lambda \}=\{3.8,0.2\}$ in Figure \ref%
{fig-diagrams-composite}c.

\textit{Meeting of second-order transition and unconventional TPT:} The
composite multiple points addressed above are located at the conventional
TPT boundaries which are in principle of first order with gap closing.
Another two composite quadruple points we did not stress are the crossing
points of the boundary $g_{c}^{\lambda ,\chi }$ (dotted
curves) and the unconventional TPT boundary, around $\{g/g_{\mathrm{%
s}},\lambda \}=\{1.4,\pm 0.6\}$ as marked by the empty squares in Figure \ref{fig-diagrams-composite}d.
The critical transition at $g_{c}^{\lambda ,\chi }$ is second-order, which is softened at finite frequencies but
still has a remnant of superradiant transition in photon number.\cite{Ying-gapped-top} The unconventional TPT would be infinite-order.

\subsection{Composite Phase Diagrams: Topological Quadruple Point being $%
\langle \sigma _{x}\rangle $-Invariant Point}

Apart from the afore-mentioned composite sextuple point formed from the
non-topological quadruple point and the $\langle \sigma _{x}\rangle $%
-sign-reversal boundary, more special is another composite sextuple point
around $\{g/g_{\mathrm{s}},\lambda \}=\{3.36,0.538\}$ (with its dual point
at $\{3.36,-0.538\}$ in $\lambda <0$ regime), as marked by the dots in Figure \ref{fig-diagrams-composite}d.
This second composite sextuple
point previously was the first topological quadruple point of Figure \ref%
{fig-topo-fix-gStark}i addressed in Section \ref{Sect-Top-Quadruple} and now
we see it happens that the conventional TPT boundary, the unconventional TPT
boundary and the $\langle \sigma _{x}\rangle $-sign-reversal boundary are
all crossing at the topological quadruple point to form a sextuple-like
point.

The $\langle \sigma _{x}\rangle $-sign-reversal boundary is also vanishing-$\langle
\sigma _{x}\rangle $ boundary indicated by the bright line Figure %
\ref{fig-diagrams-composite}a. Note here the frequency is finite, while the vanishing-$%
\langle \sigma _{x}\rangle $ boundary in the low-frequency limit, $%
g_{c}^{\sigma _{x}}$ in (\ref{LambdaC-SpinX}), is plotted as the dot-dashed
line Figure \ref{fig-diagrams-composite}b. Particularly, the topological
quadruple remains invariant when the other vanishing-$\langle \sigma _{x}\rangle $ points are moving away from dot-dashed line in the variation of
frequency. Thus we find this topological quadruple point is a $\langle
\sigma _{x}\rangle $-invariant point. Moreover, it is also an
adiabatic-invariant as similarly displayed in Figure \ref%
{fig-diagrams-composite}c where the dot-dashed line is adiabatic boundary $%
g_{c}^{\zeta 2}$ in the low-frequency limit, (\ref{gC-adiabtic-2}), while
the color change around the dot-dashed line indicates the $\zeta =1$
boundary at the finite frequency.

In Figure \ref{fig-diagrams-composite}e one can also see the sextuple-like
point in a composite phase diagram of $(n_{Z}+1)P[\text{sign}(\langle \sigma
_{x}\rangle )+\frac{1}{2}]$ in the $g$-$\chi $ plane at a fixed $\omega
=0.5\Omega $ and $\lambda =0.5$. As shown by Figure \ref{fig-diagrams-composite}f, the sextuple degeneracy will be raised if it is located close to
the Stark-JC critical boundary $g_{c}^{\mathrm{JC-Stark}}$ (Equation (\ref{gc-JC-Stark}) as plotted by dashed lines in Figure \ref{fig-diagrams-composite}e,f).
Here, unlike the leading two-peak structure both before and after transition
for points away from $g_{c}^{\mathrm{JC-Stark}}$, the GS wave function is
however of one-peak structure before the transition and two-peak structure
after,\cite{Ying-2021-AQT} which leads to different vanishing-$\langle \sigma
_{x}\rangle $ points thus the dislocation of the $\langle \sigma
_{x}\rangle $ boundaries. In contrast, the topological quadruple point is more robust and still survives there despite of the breakdown of the sextuple degeneracy

\subsection{Analytic expressions of the first topological boundary and
topological quadruple point}

By adding the Stark term to the treatment on the $\underline{0}$/$\overline{1}$ transition in the polaron picture,\cite%
{Ying-2021-AQT} we can get an analytic boundary for the first conventional TPT in the leading order
\begin{eqnarray}
g_{\mathrm{T1}}^{\lambda ,\chi } &=&\frac{2\sqrt{2}}{\sqrt{\left( 1+\lambda
\right) \left[ \left( 2+\chi \right) -\lambda \left( 2-\chi \right) \right] }%
}g_{\mathrm{s}},  \label{gT1-Polaron} \\
\lambda _{\mathrm{T1}} &=&\frac{2\sqrt{1-2\left( 2-\chi \right) g_{\mathrm{s}%
}^{2}/g^{2}}+\chi }{2-\chi },  \label{LambdaT1-Polaron} \\
\chi _{\mathrm{T1}} &=&\frac{2\left[ 4-\left( 1-\lambda ^{2}\right) g^{2}/g_{%
\mathrm{s}}^{2}\right] }{\left( 1+\left\vert \lambda \right\vert \right)
^{2}g^{2}/g_{\mathrm{s}}^{2}},
\end{eqnarray}%
which provides an analytic confirmation with a direct insight about the node variation at the TPT.\cite{Ying-2021-AQT}
From exact solution\cite{Braak2011,QHChen2020} we can also get an accurate analytic boundary
\begin{eqnarray}
g_{\mathrm{T1,E}}^{\lambda ,\chi } &=&\frac{2\sqrt{1-\chi ^{2}}}{\sqrt{%
\left( 1+\chi \right) -\lambda ^{2}\left( 1-\chi \right) }}g_{\mathrm{s}},
\label{gT1-Exact} \\
\lambda _{\mathrm{T1,E}} &=&\sqrt{\left( 1+\chi \right) [\frac{1}{1-\chi }-%
\frac{4}{g^{2}/g_{\mathrm{s}}^{2}}]},  \label{LambdaT1-Exact} \\
\chi _{\mathrm{T1,E}} &=&-\frac{1+\lambda ^{2}}{8}\frac{g^{2}}{g_{\mathrm{s}%
}^{2}}+\sqrt{[1+\frac{1+\lambda ^{2}}{8}\frac{g^{2}}{g_{\mathrm{s}}^{2}}%
]^{2}-\frac{g^{2}}{2g_{\mathrm{s}}^{2}}}. \label{ChiT1-Exact}
\end{eqnarray}%
Besides recovering $g_{\mathrm{T1}}^{\lambda ,0 } =\frac{2}{\sqrt{ 1-\lambda ^2}} g_{\mathrm{s}}$ at $\chi =0$,\cite{Ying-2021-AQT}
both $g_{\mathrm{T1}}^{\lambda ,\chi }$ and $g_{\mathrm{T1,E}}^{\lambda,\chi }$ agree with the numeric results at a finite $\chi $, as indicated by the dashed lines in Figure \ref{fig-diagrams-composite}a-d, except for some discrepancy around $%
\lambda =0$ for $g_{\mathrm{T1}}^{\lambda ,\chi }$ at a large $\chi $.

Combining (\ref{gT1-Exact}-\ref{ChiT1-Exact}) and (\ref{LambdaC-SpinX}-\ref{ChiC-SpinX}), we find the analytic locations of the topological quadruple points
\begin{eqnarray}
g_{\mathrm{TQ}}^{\chi } &=&\frac{\sqrt{2}\left( 1-\chi \right) }{\sqrt{-\chi
}},\quad \lambda _{\mathrm{TQ}}^{\chi }=\pm \frac{1+\chi }{1-\chi };
\label{g-TQ-fix-Stark} \\
g_{\mathrm{TQ}}^{\lambda } &=&\frac{2\sqrt{2}}{\sqrt{1-\lambda ^{2}}},\quad
\chi _{\mathrm{TQ}}^{\lambda }=-\frac{1-\left\vert \lambda \right\vert }{%
1+\left\vert \lambda \right\vert }.  \label{g-TQ-fix-Lambda}
\end{eqnarray}
under a given Stark coupling and under a fixed anisotropy ratio respectively, which are plotted as dots and coincide with numerics in Figures \ref{fig-topo-fix-gStark},\ref{fig-diagrams-composite}.

\subsection{Topological Quadruple Points in Large $\lambda $}

So far we have focused on $\left\vert \lambda \right\vert \leqslant 1$
regime, while topological quadruple points can also emerge in large-$\lambda
$ regime. \textbf{Figure} \ref{fig-large-lambda}a,b display the phase
diagrams of $n_{Z}$, together with $P$ represented by overlines and
underlines, respectively under a given Stark coupling ratio $\chi =0.2$ (a)
and at a fixed anisotropy strength $\lambda =2.0$ (b). For a confirmation
and a more direct view, the parity is also explicitly plotted in Figure \ref%
{fig-large-lambda}c at $\lambda =2.0$. The conventional TPTs occur between
phases $\underline{n_{Z}}$ and $\overline{n_{Z}\pm 1}$, while the
unconventional ones lie on the boundaries between phases $\overline{n_{Z}}$
and $\overline{n_{Z}\pm 1}$ or between \underline{$n_{Z}$} and $\underline{%
n_{Z}\pm 1}.$ As one sees from Figure \ref{fig-large-lambda}a the
conventional TPT boundaries remain almost unmoving in adding the Stark
coupling as compared with the $\chi =0$ boundary (dotted line). This
scenario is confirmed by Figure \ref{fig-large-lambda}b where the
conventional TPTs are not much affected in the vicinity of $\chi =0$, unless a large
amplitude of $\chi $ is involved. In a strong contrast, the unconventional
TPTs ($\underline{0}$/\underline{$1$} boundary) are very sensitive to the
variation of $\chi $, as one compares with the dashed line which represents
the unconventional TPT boundary at $\chi =0$. In the absence of the Stark
coupling, the conventional and unconventional TPTs do not cross each other.%
\cite{Ying-gapped-top} Now in adding the Stark coupling, the slow motion of
the conventional TPTs and the fast moving of the unconventional TPTs result
in the boundary crossing and thus bring about the topological quadruple
points.

\section{Mechanisms}

\label{Sect-Mechanisms}

To get an understanding for some key features of the different TPTs and the
topological quadruple points, in \textbf{Figure} \ref{fig-wavefunction}\ we
show the profiles of $\psi _{+}\left( x\right) ,$ $\psi _{-}\left( x\right) $
and $\partial _{x}\psi _{-}$, $\hat{p}^{2}\psi _{-}$, $\hat{x}^{2}\psi _{-}$
in $x$ space for GSs. They contribute to the tunneling and different interacting parts in the GS energy%
\begin{eqnarray}
E_{\Omega } &=&\frac{\Omega }{2}\int \psi _{+}\left( x\right) \psi
_{-}\left( x\right) dx, \\
E_{gy} &=&\sqrt{2}\left( -g_{y}\right) \int \psi _{+}\left( x\right)
\partial _{x}\psi _{-}\left( x\right) dx, \\
E_{p^{2}} &=&\frac{\chi \omega }{2}\int \psi _{+}\left( x\right) \hat{p}^{2}\psi
_{-}\left( x\right) dx, \\
E_{x^{2}} &=&\frac{\chi \omega }{2}\int \psi _{+}\left( x\right) \hat{x}%
^{2}\psi _{-}\left( x\right) dx,
\end{eqnarray}%
which involve subtle competitions.

\subsection{TPTs and quadruple points in $\lambda >1$\textit{\ regime%
}}

\textit{Node from infinity}: \textbf{Figure} \ref{fig-wavefunction}a-c illustrate
some typical points of different phases $\underline{0}$, $\underline{1}$, $\overline{2}$ in Figure \ref{fig-large-lambda} in $\lambda >1$ regime with $\chi
=0.1$. Around the isotropic line as in Figure \ref{fig-wavefunction}a with $%
\lambda =1.1$ and $g=2.6g_{\mathrm{s}}$, the amplitude of $g_{y}=\left(
1-\lambda \right) g/2$ is small so that $E_{\Omega }$ plays a more dominant
role which favors a nodeless state with $n_{Z}=0$ which has opposite signs
of $\psi _{+}\left( x\right) $ and $\psi _{-}\left( x\right) $ in all
positions. In a larger $\lambda $ as in Figure \ref{fig-wavefunction}b with $%
\lambda =2.0$ and $g=2.6g_{\mathrm{s}}$, note $\left( -g_{y}\right) $ is
positive here which also favors opposite signs of $\partial _{x}\psi _{-}$
and $\psi _{+}\left( x\right) $. The larger contribution of $E_{gy}$ leads
to the negative-peak replacement of $\psi _{-}$ by $\partial _{x}\psi _{-}$
in alignment (as indicated by vertical dashed line) with the positive peak of $\psi _{+}\left( x\right) $ on the
left side to get a lower energy. A node introduction from infinity will not
only enhance the negative peak of $\partial _{x}\psi _{-}$ on the left but
also make the tails of $\partial _{x}\psi _{-}$ and $\psi _{+}\left(
x\right) $ opposite in sign on the right. This energy competition creates a
node of $\psi _{\pm }\left( x\right) $\ around $x=\mp 5.2$, with $n_{Z}=1$.
This node transition between cases (a) and (b) occurs without gap closing, being an
unconventional TPT.\cite{Ying-gapped-top}

\textit{Unconventional TPT in }$\lambda >1$\textit{\ regime sensitive to }$%
\chi $: In the nonlinear Stark parts, $\hat{p}^{2}\psi
_{-}$ is oscillating to cancel itself to a large extent, thus the main
contribution lies in $\hat{x}^{2}\psi _{-}$ (which is also a reason why it
is $x$-type in $\lambda >0$ regime). Note, with a node from infinity, the tails of $\hat{x}%
^{2}\psi _{-}$ has a same sign as $\psi _{+}\left( x\right)$ as in Figure \ref{fig-wavefunction}b, which is
unfavorable for $E_{x^{2}}$ with a positive $\chi $. In this sense, $\hat{x}^{2}\psi _{-}$ is
counteracting with $\partial _{x}\psi _{-}$ in such a node introduction.
Consequently, one needs a larger $\lambda $ to strengthen the $\left(
-g_{y}\right) $ term $E_{gy}$ to trigger the unconventional TPT. This accounts for
the far boundary moving of the unconventional TPT from around $\lambda =0$
(dashed line in Figure \ref{fig-large-lambda}a) in the absence of the Stark
coupling to a larger $\lambda $ ($\underline{0}$/$\underline{1}$ boundary around $\lambda =1.7$ in Figure \ref%
{fig-large-lambda}a) in the presence of a positive Stark coupling $\chi $.

\textit{Conventional TPT in }$\lambda >1$\textit{\ regime unaffected by }$\chi $%
: On the other hand, for the conventional TPT, Figure \ref%
{fig-wavefunction}c shows the state at $\lambda =2.0$ and a larger linear
coupling $g=3.3g_{\mathrm{s}}$ after the $\underline{1}$/$\overline{2}$ transition from state in
Figure \ref{fig-wavefunction}b. Such a conventional TPT
introduces a node around the origin $x=0$, thus accompanied with gap closing
and parity reversal. In such a situation the leading variation lies around
the origin while the farther parts remain little affected. Note that $\hat{x}%
^{2}\psi _{-}$ and $\hat{p}^{2}\psi _{-}$ have similar decreasing amplitudes
but opposite signs around the origin both before and after the transition,
which leads to a cancellation effect. As a result, $\hat{x}^{2}\psi _{-}$
and $\hat{p}^{2}\psi _{-}$ together do not play much role in this
conventional TPT, unless one increases $\chi $ much to multiply their
difference. This explains the little moving of the conventional TPT
boundaries in the variations of Stark coupling as in Figure \ref%
{fig-large-lambda} (dotted line and $\underline{1}$/$\overline{2}$ boundary).

\textit{Topological quadruple points}: Since the conventional TPTs
keeps almost unmoved while the unconventional TPT is sensitive to the introduction
of the Stark coupling, their boundary meeting naturally occurs. The final
boundary crossing gives rise to the topological quadruple points.

\subsection{TPTs and quadruple points in $\lambda <1$\textit{\ regime%
}}

\textit{Unconventional TPT in }$\lambda <1$\textit{\ regime with negative }$%
\chi $: Now we look at the $\lambda <1$ regime with a
negative $\chi $, as in Figure \ref{fig-topo-fix-gStark}i. The nodeless
state ($n_{Z}=0$) in a small $g$ is similar to Figure \ref{fig-wavefunction}%
a with peak alignment of $\psi _{+}\left( x\right) $ and $\psi _{-}\left(
x\right) $ due to the dominant $E_{\Omega }$. Figure \ref{fig-wavefunction}e
shows a nodal $\underline{1}$ state in Figure \ref{fig-topo-fix-gStark}i  after the
$\underline{0}$/$\underline{1}$ unconventional TPT. Here $\left( -g_{y}\right) $ is negative, different
signs of $\partial _x\psi _{-}\left( x\right) $ and $\psi _{+}\left( x\right) $ are
unfavorable for lowering the energy of $E_{gy}$. A node introduced from
infinity as in Figure \ref{fig-wavefunction}e would not only bring wave-packet tails with same signs of
$\psi _{-}\left( x\right) $ and $\psi _{+}\left( x\right) $ to increase $E_\Omega$ but also lead to
larger tails of $\partial _x\psi _{-}\left( x\right) $ and $\psi _{+}\left( x\right) $ with different signs on the right
than the tails with same signs on the left to raise $E_{gy}$,
so there is no unconventional TPT in the absence of Stark coupling.
However, in the presence of a negative $\chi $, on both sides $\hat{x}%
^{2}\psi _{-}$ has tails with same signs as $\psi _{+}$, as in Figure \ref{fig-wavefunction}e, which reduces the
energy from $E_{x^2}$ and makes the unconventional TPT possible. Therefore, the
unconventional TPT boundary moves from $\lambda >1$ regime to $\lambda <1$
regime as in Figure \ref{fig-topo-fix-gStark}i.

\textit{Conventional TPT in }$\lambda <1$\textit{\ regime depending on }$%
\chi $: In contrast to the $\chi$-insensitiveness in $\lambda >1$ regime the
conventional TPT in $\lambda <1$ regime depends much on $\chi $ as shown in
Figures \ref{fig-topo-fix-gStark},\ref{fig-topo-fix-lambda}. Figure \ref%
{fig-wavefunction}f shows a state in the $\overline{1}$ phase of Figure \ref{fig-topo-fix-gStark}i.
Comparing with Figure \ref{fig-wavefunction}c
one sees there is no afore-mentioned cancellation effect of $\hat{x}^{2}\psi
_{-}$ and $\hat{p}^{2}\psi _{-}$ around the origin. This is because the distance of left and right
wavepackets depends on $g_{z}=\left( 1+\lambda \right) g/2$ which is much
smaller in $\lambda <1$ regime so that there is more overlap between left
and right wavepackets. Since the conventional TPT comes from the node number
variation around the origin, the transition boundary is then much influenced
by the Stark coupling with the enlarged difference of $\hat{x}^{2}\psi _{-}$
and $\hat{p}^{2}\psi _{-}$.

\textit{Invariant point}: Actually Figure \ref{fig-wavefunction}%
e,f take the points along the boundary where $\langle \sigma _{x}\rangle $
vanishes and changes the sign in Figure \ref{fig-diagrams-composite}a,b. In
these cases the wavefunction is finite in amplitude on both sides, while the
vanishing of $\langle \sigma _{x}\rangle $ comes from the cancellation
between same-sign and opposite-sign parts of $\psi _{+}\left( x\right) $ and
$\psi _{-}\left( x\right) $ with a certain position of the node. Such a
cancellation depends on the frequency $\omega $ since the size of
wavepackets will vary with the frequency \cite{Ying2020-nonlinear-bias} to
affect the cancellation situation. In contrast, the status of the
topological quadruple point is distinctive, as one side of wavepacket is
completely flat as demonstrated by Figure \ref{fig-wavefunction}d. The
vanishing of $\langle \sigma _{x}\rangle $ at the topological quadruple
point results from the vanishing local product of $\psi _{+}\left( x\right) $ and $%
\psi _{-}\left( x\right) $ rather than the cancellation. In such a
situation, other terms $\partial _{x}\psi _{-}$, $\hat{p}^{2}\psi _{-}$, $%
\hat{x}^{2}\psi $ do not come to effect either. Thus, the GS
effectively behaves like a non-interacting particle in displaced harmonic
potential ($v_{\sigma _{z}}\left( x\right) $ in (\ref{Hx})), with the
particle location adiabatically being the potential bottom position, which
is the reason why here also $\zeta =1$. Note such a status effectively being the
GS of a displaced harmonic potential remains the same for different
frequencies, this topological quadruple point appears as an invariant point
in the sense the vanishing value of $\langle \sigma _{x}\rangle $ and
adiabatic value $\zeta =1$ remain unchanged when the frequency is varying.

\section{Conclusions and discussions}

\label{Sect-Conclusions}

We have investigated the critical universality and topological
universality in light-matter interactions via a thorough study on the
first excitation gap and the GS of the QRM generally in the presence
of interaction anisotropy and nonlinear Stark coupling.

In the low-frequency limit, we have obtained both numerically and analytically all
phase boundaries of the QPTs in the GS as
well as the adiabatic boundaries and the vanishing-$\langle \sigma _{x}\rangle $ boundaries. We have extracted various scaling relations in which
physical properties collapse into the same line, respectively for different
anisotropy ratios under finite Stark coupling and variations of
both anisotropy and Stark coupling, locally around the QPTs
or globally for all coupling regions after the transitions. These scaling
relations form different levels of critical universalities. It may be worthy to mention that usually critical universality
concerns a same critical exponent around the transition while same coefficients are not required.\cite{Irish2017} Here, the scaling relations with same-line collapsing and more global range provide a stricter universality in some sense.

At finite frequencies, the critical universality breaks down and the
diversity comes to dominate. Amidst the diversity we have extracted the
topological classifications which form a new universality essentially
different from the critical universality. The critical universality
involves the second-order transitions, while the topological universality here
classifies the phases in the emerging first-order transitions for the conventional
TPTs with gap closing or the infinite-order transitions for the unconventional
TPTs without gap closing. Moreover, the universality-diversity-universality process
demonstrates that although universality and diversity are antagonists by
nature, counter-intuitively they can acquire coexistence and mutual support. We stress that both
the critical universality and the topological classification hold not only
for the linear interaction but also in the presence of nonlinear Stark
coupling, thus yielding a more robust scenario of universalities.

While the conventional TPTs and the unconventional TPTs never meet in linear
QRM,\cite{Ying-gapped-top} the presence of the nonlinear coupling enables boundary crossings of
the conventional and unconventional TPTs, which brings about the appearance of
topological quadruple points. The composite phase diagrams in combination
with the vanishing-$\langle \sigma _{x}\rangle $ and adiabatic boundaries further display the
multicriticality, composite quadruple points and composite hexaple points.
In particular, we reveal that the topological quadruple points in the
intermediate anisotropy regime are in fact $\langle \sigma _{x}\rangle $-invariant points
and adiabatically-invariant points in varying the frequency. This indicates that the locations of
such topological quadruple points can be detected by
invariant spin-flipping or tunneling points when one tunes the frequency.

Our phase diagrams and sensitivity analysis with respect to the nonlinear Stark coupling demonstrate that in addition to the anisotropy the nonlinear coupling provides another approach to manipulate both the critical QPTs and the TPTs. Especially, the unconventional TPTs are quite sensitive in response to the nonlinear coupling.

Experimentally in superconducting circuit systems\cite{flux-qubit-Mooij-1999,Bertet2005mixedModel,you024532} with deep-strong couplings\cite%
{Diaz2019RevModPhy,Wallraff2004,Gunter2009,Niemczyk2010,Peropadre2010,FornDiaz2017, Forn-Diaz2010,Scalari2012,Xiang2013,Yoshihara2017NatPhys,Kockum2017,Ulstrong-JC-1,Ulstrong-JC-2} the effective position $x$ and momentum $p$ are realistically the flux and charge of Josephson junctions and the spin can be also implemented by flux qubit, the nodal status might be detected by interference devices and magnetometer.\cite{you024532} In practice, the interaction anisotropy is highly tunable \cite{PRX-Xie-Anistropy,Forn-Diaz2010,Yimin2018} and the nonlinear Stark coupling can also be realized with adjustable amplitude and sign,\cite{Eckle-2017JPA,Stark-Grimsmo2013,Stark-Grimsmo2014,Stark-Cong2020} which could provide feasible platforms for possible tests or potential applications of our results.

\section*{Acknowledgements}

This work was supported by the National Natural Science Foundation of China
(Grant No. 11974151).

\end{document}